\documentclass[aps, prb, notitlepage, citeautoscript, superscriptaddress, eqsecnum, reprint]{revtex4-2}

\usepackage{xr}
\usepackage{amsmath}
\usepackage{amsfonts, amssymb,amsxtra}
\usepackage[]{graphicx}
\usepackage{grffile}
\usepackage{amsfonts}
\usepackage{framed}
\usepackage{bbm}
\usepackage{braket}
\usepackage{xcolor}
\usepackage{comment}

\definecolor{forestgreen}{rgb}{0.13, 0.55, 0.13}

\definecolor{tealblue}{rgb}{0.21, 0.46, 0.53}

\definecolor{red}{rgb}{1, 0, 0}

\newcommand{\av}[1]{\langle #1 \rangle}

\usepackage[colorlinks=true]{hyperref}
\hypersetup{
    unicode=false,          
    pdftoolbar=true,        
    pdfmenubar=true,        
    pdffitwindow=false,     
    pdfstartview={FitH},    
    pdftitle={},    
    pdfauthor={},     
    pdfsubject={},   
    pdfcreator={},   
    pdfproducer={}, 
    pdfkeywords={} {} {}, 
    pdfnewwindow=true,      
    colorlinks=true,       
    linkcolor=magenta, 
    citecolor=blue,        
    filecolor=magenta,      
    urlcolor=blue           
}

\begin{document}
\title{Emergent Potts Order in a Coupled Hexatic-Nematic XY model}

\author{Victor Drouin-Touchette}
\affiliation{Center for Materials Theory, Rutgers University, Piscataway, New Jersey 08854, USA}

\author{Peter P.~Orth}
\email{porth@iastate.edu}
\affiliation{Ames Laboratory, Ames, Iowa 50011, USA}
\affiliation{Department of Physics and Astronomy, Iowa State University, Ames, Iowa 50011, USA}

\author{Piers Coleman}
\affiliation{Center for Materials Theory, Rutgers University, Piscataway, New Jersey 08854, USA}
\affiliation{Hubbard Theory Consortium and Department of Physics, Royal Holloway, University of London, Egham, Surrey TW20 0EX, UK}

\author{Premala Chandra}
\affiliation{Center for Materials Theory, Rutgers University, Piscataway, New Jersey 08854, USA}

\author{Tom C. Lubensky}
\affiliation{Department of Physics and Astronomy, University of Pennsylvania,
209 South 33rd St, Philadelphia, PA 19104}

\begin{abstract}
Addressing the nature of an unexpected smectic-A' phase in liquid crystal 54COOBC films, we perform large scale Monte Carlo simulations of a coupled hexatic-nematic XY model. The resulting finite-temperature phase diagram reveals 
a small region with composite Potts $\mathbb{Z}_3$ order above the
vortex binding transition; this phase is characterized by relative
hexatic-nematic ordering though both variables are disordered. 
The system develops algebraic hexatic and nematic order 
only at a lower temperature. This 
multi-step melting scenario agrees well with the experimental observations of a sharp specific heat anomaly that emerges above the onset of hexatic positional order. We therefore propose that the smectic-A' phase is characterized by composite Potts order and bound-states of fractional vortices.
\end{abstract}
\date{\today}
\maketitle

\section{Introduction} 
\label{sec:introduction}

%

\begin{figure}
    \centering
    \includegraphics[width=\linewidth]{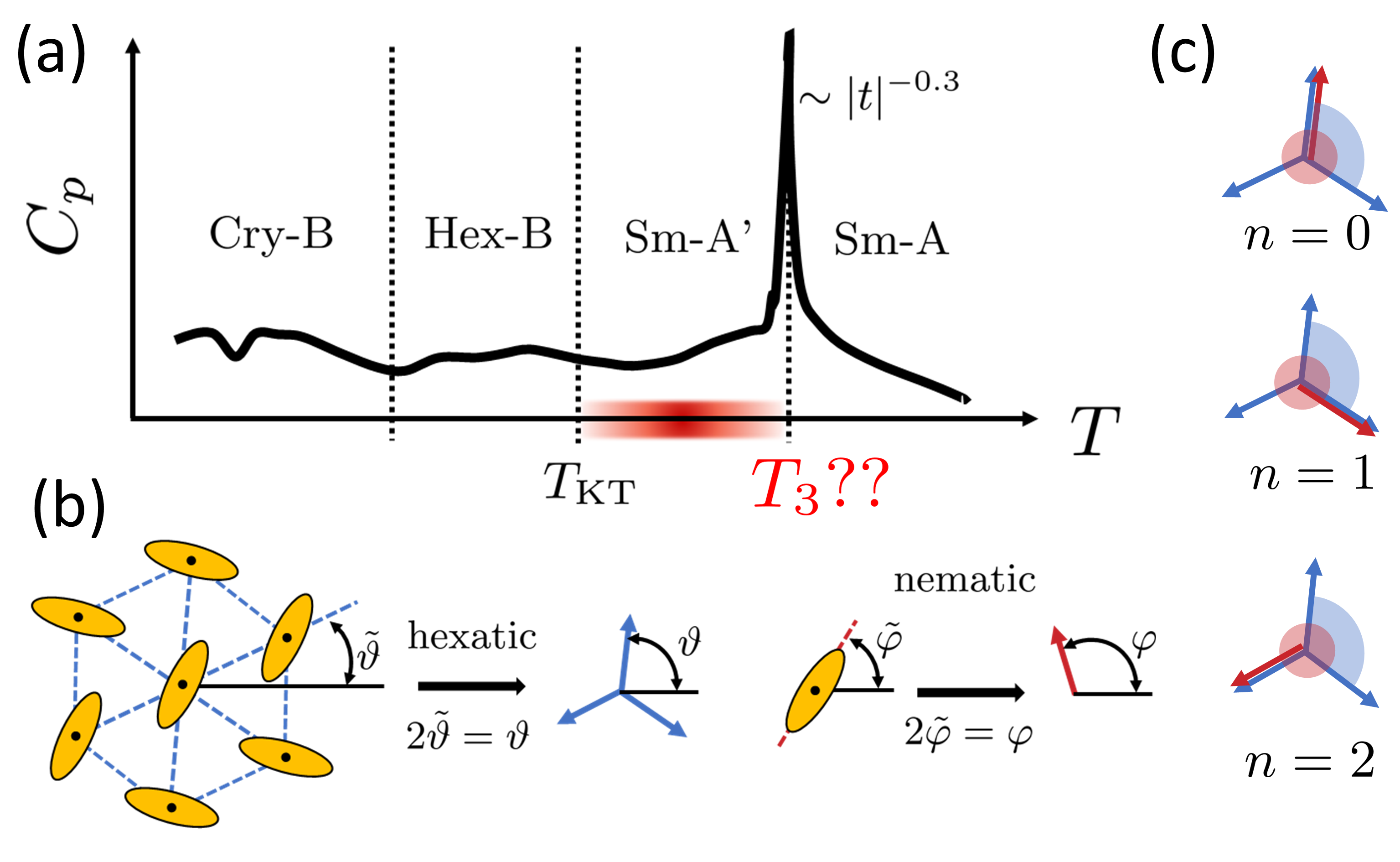}
    \caption{(a) Schematic of the specific heat curve obtained by Chou \textit{et al}~\cite{chou1998multiple} upon melting of 54COOBC. (b) The hexatic bond-orientational $\vartheta$ and molecular nematic $\varphi$ degrees of freedom in the liquid crystal film and their representations in the generalized XY model of (\ref{eq:1.2-model_xy}) (c) Three relative configurations of $\tilde\sigma = 2\pi n /3 = \vartheta - \varphi$ (as presented in Eq.~\ref{pottsOP}) for the angular variables in (\ref{eq:1.2-model_xy}). Shaded areas represent the domain of their respective variables.
    }
    \label{fig:figure_intro}
\end{figure}

The appearance of a sharp specific heat signal in the multiple-step
melting sequence \cite{jin1996nature, chou1997electron,chou1998multiple} of
certain free-standing liquid crystal films (Fig.~\ref{fig:figure_intro} (a)) remains an outstanding 
mystery. Such transitions are typically  associated 
with the unbinding of topological defects~\cite{Berezinskii,KTordering,KTordering2} that are not associated with
acute thermodynamic signatures.   
Here we revisit this unsolved 
problem, bringing to it modern concepts and methods that have been developed 
in other areas. 
In a minimalist model~\cite{bruinsma1982hexatic, aeppli1984hexatic,jiangMonteCarloSimulation1993,jiang1996monte}, we explore whether vortex fractionalization
can lead to an emergent three-state Potts phase above a 
Kosterlitz-Thouless binding
temperature. Supported by analytical arguments, we perform this study using large-scale Monte Carlo simulations, 
finding a parameter region of the phase diagram consistent with experimental observations. 

Optical reflectivity, electron diffraction and specific heat measurements on
free-standing two-dimensional (2D) films of 54COOBC ($n$-pentyl-$4'$-$n$-pentanoyloxybiphenyl-$4$-carboxylate)~\cite{jin1996nature, chou1997electron,chou1998multiple} provide the experimental
motivation for our work. Theoretically, a two-stage melting sequence was
expected in these 2D films with an intermediate hexatic phase residing
between the isotropic liquid and the 2D crystalline solid. By contrast, experimentalists observed three-step 
melting with {\sl two} phases separating the solid and the liquid states.  
In particular
they detected a hexatic phase and then at higher temperatures a mystery
intermediate liquid phase with no long-range orientational order. 
The experimentalists referred to this unexpected  ``hidden order'' (HO)  
in 2D films of 54COOBC as 
smectic-A$'$ (Sm-A$'$) phase. It is sandwiched between a (disordered) 
isotropic smectic-A (Sm-A) phase at higher temperature and a hexatic (Hex-B) 
phase with bond orientational quasi-long-range order (QLRO) at 
lower temperature 
(see  Fig.~\ref{fig:figure_intro} (a)). 
At even lower temperatures, the system develops positional 
QLRO in the hexagonal crystalline phase (Cry-B). 
The transition from disordered Sm-A into HO Sm-A$'$ phase is characterized 
by a pronounced specific heat anomaly that is in striking disagreement 
with the broad features predicted by the 2D melting theory of 
Kosterlitz, Thouless, Halperin, Nelson and Young (KTHNY) \cite{Berezinskii,KTordering,KTordering2,joseRenormalizationVorticesSymmetrybreaking1977,halperin1978theory, HalperinNelson, young1979melting}. 
The scaling exponent associated with
this specific heat was reported to be
$\alpha_{\text{54COOBC}} = 0.30 \pm 0.07$~\cite{jin1996nature}, suggesting
that the HO Sm-A$'$ phase has 3-state Potts order at temperatures
above the conventional vortex binding transition.


\begin{figure*}
    \centering
    \includegraphics[width=0.9\linewidth]{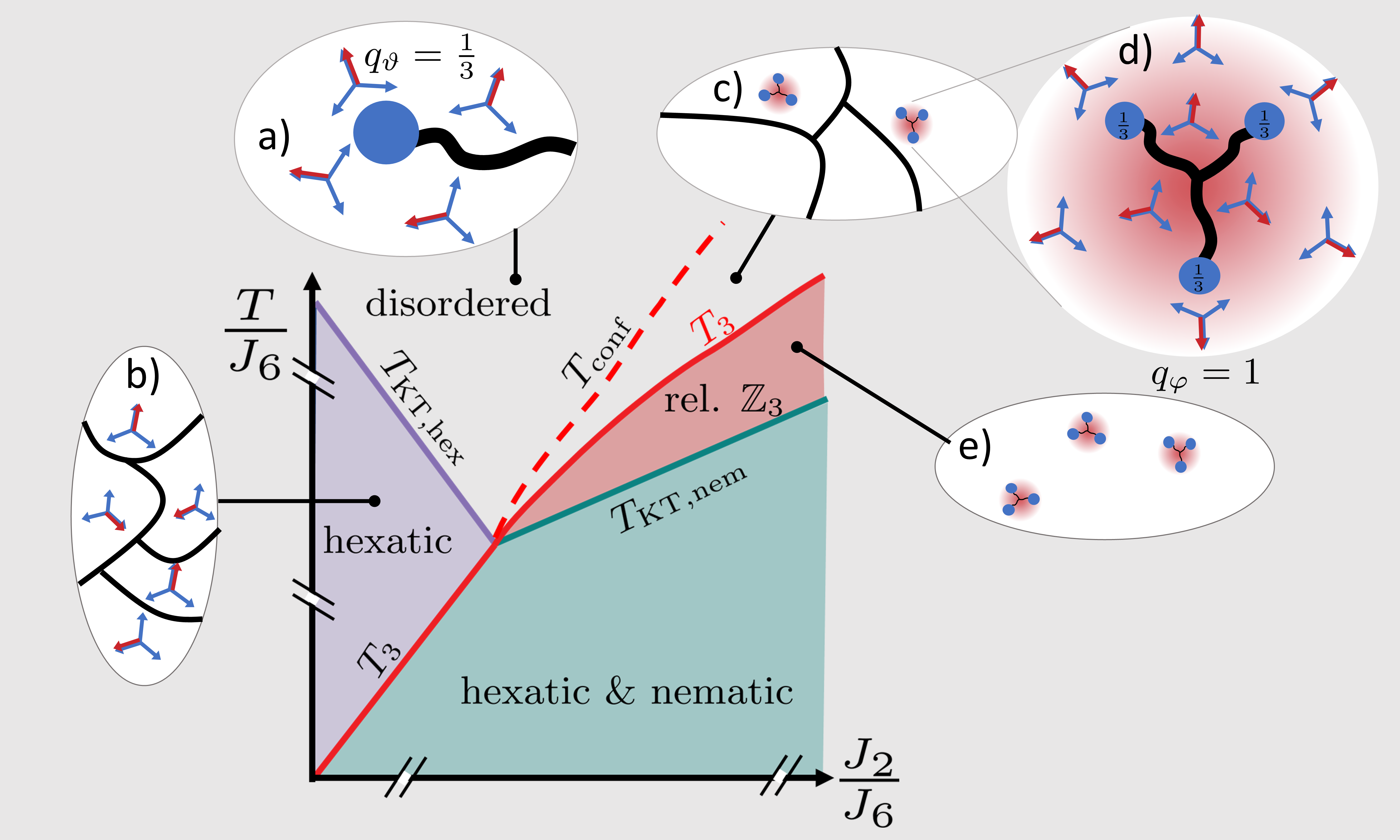}
    \caption{Schematic phase diagram of the coupled ($\lambda' \neq 0$)
generalized XY model (\ref{eq:1.2-model_xy}) (center), where 
$T_{\rm KT, hex}$ and $T_{\rm KT, nem}$ are the binding temperatures 
of $q_\vartheta = 1/3$ (blue tri-arrows denote the hexatic $\vartheta$ variable) and
$q_\varphi = 1$ (red arrows denote the nematic $\varphi$ variable) vortices, respectively. Here $T_3$ refers to a three-state
Potts transition that is well-established at temperatures below $T_{\rm KT,hex}$, but could persist above $T_{\rm KT,nem}$. The disordered phase is characterized in (a) by free hexatic vortices (blue dot), which correspond to fractional nematic vortices, where $q_\varphi = 1/3$ due to the presence of a Potts domain wall (in black) where $\Delta n = \pm 1$. For $J_2 \ll J_6$, these hexatic vortice then become bound in neutral pairs as one lowers the temperature below $T_{\rm KT,hex}$. The hexatic ordered phase is presented pictorially in (b), where the hexatic variable (blue) is still ordered, but the lack of a single Potts domain (i.e. presence of Potts domain walls where $\Delta n = \pm 1$) prevents order in the $\varphi$ (red). Only below $T_3$ does a single domain cover the whole system. As one increases $J_2/J_6$, a new sequence of phase transition occurs. Firstly, there is the development of composite $q_\varphi = 1$ vortices, bound states of 
three $q_\vartheta=1/3$ defects,
at a confinement transition labeled as $T_{\rm conf}$. Below this, there is a network of
local Potts domains (c) due to the absence of free $q_\vartheta = 1/3$ defects
and their associated ``dangling'' Potts domain walls. Within these domains, neither $\varphi$ nor $\vartheta$ is ordered, and a composite nematic vortex can sit there. A zoom of such vortex is presented in (d). Beyond the red halo corresponding to the vortex core, the vortex is perceived as one with $q_\varphi = 1$ and $q_\theta = 1$. Within this core, however, is a structure where three $q_\vartheta = 1/3$ vortices are bound together by the Potts domain walls joining them. Below $T_{3}$, the system becomes a single domain of the local Potts order (e), and there are free composite nematic vortices. 
Motivated by experiment, we confirm the presence of a Potts phase at a 
temperature $T_3 > T_{\rm KT,nem}$ 
associated with relative ordering of the hexatic and nematic degrees
of freedom and the vanishing of the Potts domain walls.
}
    \label{fig:figure_intro2}
\end{figure*}

Theoretically this problem can be studied by a minimalist coupled
hexatic-nematic model on a 2D square lattice~\cite{bruinsma1982hexatic, aeppli1984hexatic,jiangMonteCarloSimulation1993,jiang1996monte}, 
where the hexatic degrees of freedom ($\tilde\vartheta$, invariant under $\tilde\vartheta_i \rightarrow \tilde\vartheta_i + 2 \pi n/6$ for integer $n$) describes the orientational order of neighboring molecules' center of mass, and the nematic degrees of freedom ($\tilde\varphi$, invariant under $\tilde\varphi_i \rightarrow \tilde\varphi_i + \pi n$ for integer $n$) corresponds to the orientation of the rod shaped molecules of 54COOBC with respect to a fixed laboratory axis (see Fig.~\ref{fig:figure_intro} (b)),
 
\begin{align}
\mathcal{H} & = - J_2 \sum_{\av{i,j}} \cos \bigl[ 2 (\tilde\varphi_i - \tilde\varphi_j)\bigr] - J_6 \sum_{\av{i,j}} \cos \bigl[ 6 (\tilde\vartheta_i - \tilde\vartheta_j) \bigr] \nonumber \\ & \qquad - \lambda' \sum_i \cos \bigl[ 6 (\tilde\vartheta_i - \tilde\varphi_i)\bigr] \,.
\label{eq:1.1-model_varphi}
\end{align}
The hexatic-nematic coupling $\lambda' >0 $ arises from van-der-Waals interactions between the molecules, and favors a parallel relative alignment~\cite{bruinsma1982hexatic, aeppli1984hexatic, kohandel2003hexatic}.

By rescaling $\vartheta = 2\tilde\vartheta$ and $\varphi = 2\tilde \varphi $ so that
the degrees of freedom are vectors rather than directors, we can 
reexpress Eq.~\ref{eq:1.1-model_varphi} as a generalized XY model
\begin{align}
\mathcal{H'} & = - J_2 \sum_{\av{i,j}} \cos \bigl[ (\varphi_i - \varphi_j)\bigr] - J_6 \sum_{\av{i,j}} \cos \bigl[ 3(\vartheta_i - \vartheta_j) \bigr] \nonumber \\ & \qquad - \lambda' \sum_i \cos \bigl[ 3 (\vartheta_i - \varphi_i)\bigr] \,.
\label{eq:1.2-model_xy}
\end{align}
The phase diagram of the uncoupled ($\lambda'=0$) model has nematic and
hexatic Kosterlitz-Thouless (KT) transitions.  
Above these defect-binding temperatures, free vortices are 
present with ``charge'' $q$ associated  
with the phase winding $2 \pi q$ around them; here $q_\vartheta = \frac{\Delta \vartheta}{2\pi} =\frac{1}{3}$
and $q_\varphi = \frac{\Delta\varphi}{2\pi} = 1$.

When the coupling $\lambda'$ is finite, the two vortex types are
no longer independent, since now $\vartheta - \varphi \equiv \frac{2\pi}{3}n \; (\text{mod} \;2\pi)$
where $n$ is an integer (the $\text{mod} \; 2\pi$ simply associates $\tilde{n}_i = -1, -2$ to $n_i = 2, 1$, respectively). There are therefore three inequivalent relative alignments of $\varphi$
and $\vartheta$ (see Fig.~\ref{fig:figure_intro} (c)). This suggests the presence of a well-defined 3-state Potts order parameter, which we write as
\begin{equation}\label{pottsOP} 
M_{\tilde\sigma} = \sum_i \exp{[i \tilde\sigma_i]} = \sum_i \exp{[i (\vartheta_i - \varphi_i)]} \;.
\end{equation}
Since $\tilde\sigma_i = \frac{2\pi}{3}n_i$ for finite $\lambda'$, then a finite $m_{\tilde\sigma} = \langle |M_{\tilde\sigma}| \rangle$ is the direct consequence of long-range order in the relative Potts variable.

Another drastic consequence of a finite $\lambda'$, the vortex charges are now related by the relation
\begin{equation}\label{qlambda} 
q_\vartheta \equiv q_\varphi + \frac{\Delta n}{3}  \; (\text{mod} \; 1) \quad\quad\quad (\lambda' \neq 0) \;,
\end{equation}
where $\Delta n$ is the number of walls encountered that separate different Potts domains, where $\tilde\sigma \rightarrow \tilde \sigma + \frac{2\pi}{3}$ across the Potts domain wall. This expression has two important consequences:

\begin{itemize}

\item[$\circ$] $q_\vartheta$ defects are bound to Potts domain walls 
(in black in Fig.~\ref{fig:figure_intro2} (a))
which follows from Eq.~\ref{qlambda} 
as $\frac{1}{3} = 0 + \frac{1}{3}$.

\item[$\circ$] Integer $q_\varphi=1$ vortices are composites formed of $q_\vartheta=\frac{1}{3}$
defects  bound by domain walls (Fig.~\ref{fig:figure_intro2} (e)) since, referring to Eq.~\ref{qlambda}, $3 \times \frac{1}{3} = 1$.  

\end{itemize}

When there are no free $q_\vartheta =\frac{1}{3}$ charges and thus no ``dangling'' Potts domain walls, a
unique local Potts order parameter can be defined.  Therefore, the
binding of $q_{\vartheta}=\frac{1}{3}$ vortices at the hexatic KT transition
for $J_2 \ll J_6$ results in a network of Potts walls 
(Fig.~\ref{fig:figure_intro2} (b)) separating
distinct local Potts domains.  At some lower temperature one expects
one such domain to dominate the system leading to long-range
Potts order, and a finite value of the order parameter $m_{\tilde \sigma}$ from Eq.~\ref{pottsOP}.  Indeed previous computational work in this parameter
regime confirms the presence of a Potts 
ordered phase below the hexatic 
Kosterlitz-Thouless 
transition~\cite{jiangMonteCarloSimulation1993,jiang1996monte}
as indicated in the schematic phase diagram in 
Figure~\ref{fig:figure_intro2}.

The situation on the nematic KT side of the phase diagram ($J_2 \gg J_6$)
is more subtle.  From previous studies \cite{nelson1980solid, dierker1986consequences, radzihovsky2008superfluidity,fellowsUnbindingGiantVortices2012, nitta2012baryonic,shiBosonPairingUnusual2011,serna2017deconfinement,kobayashi2020z, kobayashi2020vortex, kobayashi2019berezinskii}, we expect that
there exists a parameter regime where bound states of three $q_\vartheta$ 
vortices form above the nematic KT transition (Fig.~\ref{fig:figure_intro2} (d)). This is driven by confinement of $q_\vartheta = 1/3$ fractional vortices into a composite and extended $q_\varphi = 1$ vortex.  At this ``confinement''
temperature, local Potts order develops. If this scale was merely a crossover, then there could not be any lower temperature transition into Potts long-range order. By continuity with the $J_2 \ll J_6$ side, the low-temperature ordered phase must have Potts LRO. Therefore, one has to embrace the results that the confinement is a true phase transition \cite{serna2017deconfinement}. For lower temperatures than the confinement transition, the associated Potts domain walls disappear and
there is long-range Potts order (Fig.~\ref{fig:figure_intro2} (e)), again reflected in a finite value of the order parameter $m_{\tilde \sigma}$. This will certainly be the case
when the composite vortices bind at $T_{\rm KT, nem}$;  but the coincidence of these two
transition temperatures would surely be indicative of an underlying
unknown symmetry.  
The 54COOBC measurements suggest Potts 
ordering at temperatures above that of the nematic vortex
binding~\cite{jin1996nature, chou1997electron,chou1998multiple}. 
We therefore probe whether we can tune Eq.~\ref{eq:1.2-model_xy} 
to a parameter regime of its phase diagram
where there is a Potts phase above the nematic 
Kosterlitz-Thouless transition (Fig. \ref{fig:figure_intro2} (c), on the right side); this would emulate the experimental 
observation.

The proposed Potts phase involves the relative orientation of
the disordered hexatic and the nematic phases. It is then
composite in nature since it will necessarily involve higher-order
correlations of the primary hexatic and nematic angles.  
A simple example of such composite order is nematicity in 
localized spin models such as the 
two-dimensional (2D) $J_1$-$J_2$ Heisenberg square lattice 
model~\cite{chandra1990ising,weberIsingTransitionDriven2003,capriottiIsingTransitionTwoDimensional2004}. There, nematic order corresponds to a relative ordering of spins, which breaks fourfold lattice rotation symmetry even though the 
spins retain their full SU(2) rotation symmetry.  A path towards 
the development of composite order is via partial melting of an underlying multi-component primary order parameter due to increasing fluctuations.
Being intertwined with a primary order, composite orders often occur in proximity to other ordered phases, giving a natural explanation for the complexity of phase diagrams observed in correlated systems. Examples are frustrated magnets~\cite{andreevSpinNematics1984, chandra1990ising, weberIsingTransitionDriven2003,capriottiIsingTransitionTwoDimensional2004, nakatsuji2005,mulderSpiralOrderDisorder2010,zhitomirskyOctupolarOrderingClassical2008,henleyCoulombPhaseFrustrated2010, orthEmergentCriticalPhase2012,chernDipolarOrderDisorder2013,rosalesBrokenDiscreteSymmetries2013,orthEmergentCriticalityFriedan2014,jeevanesan2015emergent}, unconventional superconductors~\cite{fangTheoryElectronNematic2008,xuIsingSpinOrders2008,fernandes2010effects,fradkinNematicFermiFluids2010,herlandPhaseTransitionsThree2010,fernandesIntertwinedVestigialOrder2019,venderbosPairingStatesSpin2018,jiangChargeOrderBroken2019,agterbergPhysicsPairdensityWaves2020,cao2020nematicity}, ultracold atoms~\cite{gopalakrishnanIntertwinedVestigialOrder2017} and liquid crystals~\cite{lubenskyTheoryCriticalPoint1996}.

Above the defect-binding transitions, vortices are the 
``elementary excitations'' of the hexatic-nematic coupled
system of Eq.~\ref{eq:1.2-model_xy}
with charge associated
with their phase winding.  It is known that, as is the
case here,  integer 
charge vortices can lower their energy by splitting into multiple
fractional charge vortices linked by domain wall 
strings \cite{radzihovsky2008superfluidity}, if it is energetically favored.
Such fractionalization
describes the phenomenon where the elementary excitations of a system carry fractional charges (or quantum numbers) that are different from ones of the microscopic degrees of freedom~\cite{wilczekQuantumMechanicsFractionalspin1982}. 
Fractional excitations are often confined into objects with integer charge by a strong string tension force, yet can unbind at a 
confinement-deconfinement transition~\cite{polyakov1977quark, fradkin1979phase, senthil2004deconfined,serna2017deconfinement}.
Confinement is well-known from elementary particle physics, where it describes the binding of quarks, which carry fractional electric charge, into integer charge baryons or mesons. It also occurs frequently in condensed matter systems, for example, in low-dimensional magnets~\cite{tennantUnboundSpinonsAntiferromagnetic1993,mourigalFractionalSpinonExcitations2013,kitaevAnyonsExactlySolved2006}, spin ice models~\cite{castelnovo2008magnetic,henleyCoulombPhaseFrustrated2010}, quantum Hall systems~\cite{laughlinAnomalousQuantumHall1983}, coupled atomic-molecular superfluids~\cite{radzihovsky2008superfluidity}, and generalized XY models with vector magnetic or nematic degrees of freedom~\cite{lee1985strings,nitta2012baryonic,serna2017deconfinement}. Here we propose that the 54COOBC films
provide another setting for this phenomenon, arguing that confinement
of fractional nematic vortices, as shown in Fig~\ref{fig:figure_intro2} (e), leads to the composite Potts phase
at temperatures above that of the nematic Kosterlitz-Thouless transition. Confinement of fractional nematic vortices drives the binding of three elementary $q_\vartheta = 1/3$ hexatic vortices. This bound state formation removes the dangling ends of Potts domain walls and thus enables the appearance of a well-defined Potts order parameter in the system.

In this paper, we use large-scale parallel-tempering classical Monte-Carlo simulations to investigate the finite temperature phase diagram of the model described in equation~\ref{eq:1.2-model_xy}, which captures the relevant degrees of freedom in the experimental system and is consistent with its symmetries. We demonstrate that it contains a small region where Potts order exists even though the underlying hexatic and nematic angles remain disordered. We show that the transition to the $\mathbb{Z}_3$ ordered phase lies in the 2D Potts universality class, which is characterized by a specific heat divergence with scaling exponent $\alpha = 1/3$, 
in good agreement with the experimental scaling exponent. Hexatic and nematic QLRO only develop at slightly lower temperatures via a Kosterlitz-Thouless (KT) phase transition.
We support our unbiased numerical findings by analytical arguments that describe how fractionalization of nematic vortices can lead to extended vortex cores and separate the Potts transition from the KT transition. The sequence of upper Potts and lower KT transitions serve as a natural explanation of the experimentally observed melting process from Hex-B to Sm-A' to Sm-A. This resolves the long-standing puzzle of why the two-step melting theory of KTHNY, which accounts well for the experimental observations at the Hex-B to Sm-A' (and the lower Cry-B to Hex-B) transitions, fails to explain the observed features at the Sm-A' (HO) to Sm-A transition.

The remainder of the paper is organized as follows: in Sec.~\ref{sec:generalized_hexatic_nematic_xy_model}, we explore the thermodynamical phase diagram of the coupled hexatic-nematic model that describes 54COOBC films for zero and finite $\lambda'$ coupling, using analytical estimates. We highlight the interplay of vortex confinement, Potts domain-walls and fractionalization, which plays a key role.
In Sec.~\ref{sec:Monte_Carlo_algorithm}, we describe the Monte Carlo algorithm, and introduce the observables that we use to identify the different thermodynamic phases. In Sec.~\ref{sec:Monte_Carlo_results}, we present results of our Monte Carlo simulations. This includes our main result: the finite temperature phase diagram of the coupled hexatic-nematic model. We discuss the behavior of the system in different regions of the phase diagram. We close the paper in Sec.~\ref{sec:conclusion} with concluding remarks and outlook for future research directions.

\section{Hexatic-nematic XY model} 
\label{sec:generalized_hexatic_nematic_xy_model}

For technical reasons, in the rest of the paper we study a model in the same universality class as that of Eqs.~\ref{eq:1.1-model_varphi} and \ref{eq:1.2-model_xy}. This is done through a transformation of the hexatic and nematic degrees of freedom of the minimal coupled model of Eq.~\ref{eq:1.1-model_varphi}. Rescaling the angles $\theta_i = 6 \tilde\vartheta_i, \phi_i = 2 \tilde\varphi_i$ (equivalently, $\theta_i = 3 \vartheta_i, \phi_i = \varphi_i$) so they both cover the range $\theta_i, \phi_i \in [0, 2 \pi)$, yields the dimensionless expression
\begin{align}
    \mathcal{H}/J &=  - \Delta \sum_{\av{i,j}} \cos (\phi_i - \phi_j) - (2 - \Delta) \sum_{\av{i,j}} \cos (\theta_i - \theta_j)\nonumber \\ & - \lambda \sum_i \cos ( \theta_i - 3 \phi_i) \,.
\label{eq:2.1-model_phi}
\end{align}
Here, we have introduced $J = \frac12 (J_2 + J_6)$, $\lambda = \lambda'/J$ and $ \Delta \equiv J_2/J$ such that $0 \leq \Delta \leq 2$ covers all exchange coupling ratios $J_2/J_6$. A value of $\Delta = 1$ corresponds to the isotropic limit $J_2 = J_6$. This description of our minimal model in terms of two $\text{O(2)}$ variables is most useful for the Monte-Carlo study we provide in the next sections, hence the change of variables. 

\begin{figure}
    \centering
    \includegraphics[width=\linewidth]{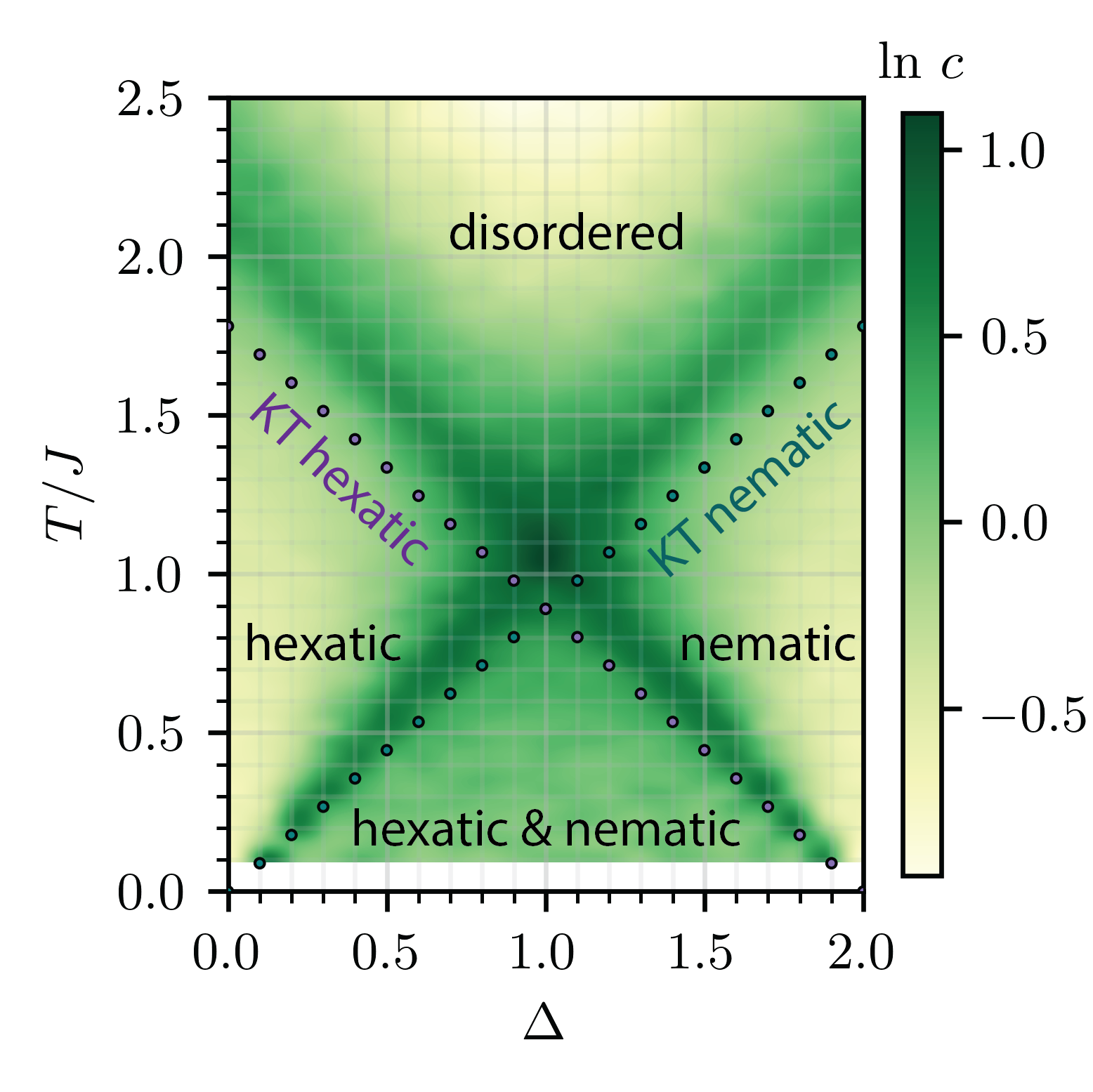}
    \caption{Numerical phase diagram of uncoupled $\lambda = 0$ hexatic-nematic XY model in Eq.~\eqref{eq:2.1-model_phi} as a function of temperature $T$ and ratio of exchange couplings $\Delta = J_2/J$, obtained from classical Monte-Carlo simulations. Background color density denotes specific heat per site $c$ at system size $L=40$ (see logarithmic color bar). We distinguish four different phases: (i) a disordered phase with purely short-range correlations, (ii) a hexatic phase with algebraic correlations of $\langle \cos(\theta_i - \theta_j) \rangle \sim |i-j|^{-\eta_6(T)}$, (iii) a nematic phase with algebraic correlations of $\langle \cos(\phi_i - \phi_j) \rangle \sim |i-j|^{-\eta_2(T)}$, and (iv) an ordered phase with algebraic correlations of both hexatic and nematic degrees of freedom. Here, $\langle \mathcal{O} \rangle$ denotes the thermal average. The KT transition temperatures (purple and blue dots) are obtained from the universal stiffness jump criterion $\rho^{(a)}(T_{a,KT}) = 2 T_{a,\text{KT}}/\pi$ for nematic, $a=2$, and hexatic, $a=6$ variable, extrapolated to infinite system size.}
    \label{fig:figure_uncoupled_PD}
\end{figure}

In the remainder of this section, we provide an intuitive and semi-analytical description of the expected phase diagram of this model. Starting from the uncoupled model is subsection~\ref{subsec:uncoupled_hexatic_nematic_model}, we first show that the hexatic-nematic coupling term is a relevant perturbation that tends to induce relative $\mathbb{Z}_3$ Potts order at temperatures larger than the KT transition temperature. Furthermore, we point out in subsection~\ref{sub:potts_transition} the important role of vortex excitations in the system, especially as to how they differ from those in the uncoupled model. In subsection~\ref{ssub:vortex_fractionalization_in_potts_ordered_phase}, we show why the fractionalization of the nematic vortices leads to extended vortex cores, and that is a necessary requirement for the Potts transition to occur above the KT transition. Finally we review some numerical results of related models in subsection~\ref{ssec:previous_studies}.

\subsection{Uncoupled model and relevance of coupling term}
\label{subsec:uncoupled_hexatic_nematic_model}
In the absence of a coupling term, $\lambda = 0$, hexatic and nematic degrees of freedom undergo separate KT transitions at temperatures~\cite{KTordering, chaikin1995principles, hasenbusch2005multicritical, hasenbusch2008binder,hsieh2013finite}
\begin{subequations}
\begin{align}
T^{(\lambda = 0)}_{2, \text{KT}}(\Delta)/J &= 0.89 \, \Delta \label{eq:2.3a}
\\
T^{(\lambda = 0)}_{6, \text{KT}} (\Delta)/J &= 0.89 \, (2 - \Delta) \,.
\label{eq:2.3b}
\end{align}
\label{eq:T_KT_at_zero_lambda}
\end{subequations}
The resulting phase diagram is shown in Fig.~\ref{fig:figure_uncoupled_PD} and exhibits four phases. For details on the classical Monte-Carlo simulations used to obtain this phase diagram, see section~\ref{sec:Monte_Carlo_algorithm}. These phases distinguish regions with short-range order (SRO) of the hexatic and nematic degrees of freedom from regions with QLRO. The phase transitions lie in the KT universality class and are thus characterized by a sudden jump of hexatic ($p=6$) or nematic ($p=2$) spin stiffness $K_p = \rho_p/T$ from zero to the universal value $K_p(T_{p, \text{KT}}) = 2/\pi$.
The background color in Fig.~\ref{fig:figure_uncoupled_PD} shows that the specific heat $c$ exhibits a broad hump above the transition at about $T = 1.1 T_{\text{KT}}$~\cite{chaikin1995principles}. In the isotropic limit, $\Delta = 1$, the two KT transitions occur at the same temperature: $T^{(\lambda = 0)}_{\text{KT}} = 0.89 J $.

In order to determine the effect of a finite hexatic-nematic coupling term $(\lambda > 0)$ on the phase diagram, we calculate its renormalization group (RG) scaling dimension $\mathcal{D}_\lambda$. This can be done straightforwardly  at $\Delta = 1$ and $T_{\text{KT}}$, where free vortex excitations are absent, to yield
\begin{align}
\mathcal{D}_\lambda = 2 - \frac{1}{4\pi} \Bigl(\frac{1}{K_{R,6}} + \frac{9}{K_{R,2}} \Bigr) = \frac34 \,.
\label{eq:2.4}
\end{align}
Here, we have used that $K_{R,p} = 2/\pi$ is the exact value of the renormalized spin stiffness at the KT transition temperature $T^{(\lambda = 0)}_{p, \text{KT}}$~\cite{chaikin1995principles}. Note that $K_{p}$ is independent of $p$ only for the rescaled Hamiltonian in Eq.~\eqref{eq:2.1-model_phi}, where the minimal phase winding of both hexatic and nematic vortices is equal to $2 \pi$, but the final result ($\mathcal{D}_\lambda = 3/4$) is identical if one uses Eq.~\eqref{eq:1.2-model_xy}. Further details of this derivation are presented in appendix~\ref{appendix1}.

A positive scaling dimension $\mathcal{D}_\lambda > 0$ indicates that $\lambda$ is a relevant perturbation at $T^{(\lambda=0)}_{\text{KT}}$ and will drive the system away from the uncoupled KT fixed point. This suggests that the system develops long-range Potts order above the KT transition, $T_{\mathbb{Z}_3} > T_{\text{KT}}^{(\lambda = 0)}$.
Certainly as $\lambda$ grows towards longer length scales, hexatic and nematic angles are forced to arrange into one of the three parallel configurations, shown in Fig.~\ref{fig:figure_intro}(d) for the equivalent $\vartheta$ and $\varphi$ variables. The constraint due to $\lambda$, as expressed for Eq.~\eqref{eq:2.1-model_phi}, is then
\begin{align}
\theta_i - 3 \phi_i = 2 \pi n_i \Rightarrow \phi_i = \frac{\theta_i}{3} + \frac{2 \pi n_i}{3} \,.
\label{eq:2.6-locking_constraint}
\end{align}

Here, $n_i = -1,0,1$ is a $\mathbb{Z}_3$ degree of freedom. Note that if the locking constraint is fulfilled, the model~\eqref{eq:2.1-model_phi} becomes equivalent to a generalized XY model~\cite{lee1985strings,korshunov1985possible}, in our case for $q=3$~\cite{romanoTopologicalTransitionsTwodimensional2006,poderosoNewOrderedPhases2011,canovaKosterlitzthoulessPottsTransitions2014}. We will review the results of the fully locked model in subsection~\ref{ssec:previous_studies}. The locking transition occurs as a crossover at a temperature $T_\lambda(\lambda)$ that depends on $\lambda$. Since $\mathcal{D}_\lambda > 0$, it follows that $T_\lambda > T_{\text{KT}}$ for all nonzero values of $\lambda$. For small initial values of $\lambda \ll 1$, $T_\lambda$ can be estimated from analyzing the RG equation of an XY model in a threefold potential, which yields $T_\lambda \approx \frac{8 \pi}{9} J$~\cite{joseRenormalizationVorticesSymmetrybreaking1977}. For large initial values of $\lambda \gg 1$, the locking occurs at a temperature $T_\lambda \gg J$. For $T < T_\lambda$, the symmetry of the model is lowered from $\text{O(2)} \times \text{O(2)}$ $\rightarrow $ $\text{O(2)} \times \mathbb{Z}_3$.

\subsection{Vortex excitations in the coupled model} 
\label{sub:potts_transition}
For $T < T_\lambda$ the relative angle between nematic and hexatic degrees of freedom is a $\mathbb{Z}_3$ variable $n_i$. Being discrete, $n_i$ can develop LR order at a finite transition temperature $T_3$. The central question is whether $T_3$ lies above or below the KT transition of the ``center-of-mass'' O(2) variable. If $T_3 > T_{\text{KT}}$ the situation corresponds to the experimentally observed order of phase transitions in 54COOBC. This turns out to be a rather delicate issue~\cite{jiangMonteCarloSimulation1993, jiang1996monte} that requires a careful and unbiased large-scale computational effort, which is described below in Secs.~\ref{sec:Monte_Carlo_algorithm} and~\ref{sec:Monte_Carlo_results}. We note that several studies of related coupled XY models, obtained by taking the limit of $\lambda \rightarrow \infty$ in Eq.~\eqref{eq:2.1-model_phi}, revealed intriguing behavior close to $T_{\text{KT}}$ at $\Delta = 1$~\cite{poderosoNewOrderedPhases2011, canovaKosterlitzthoulessPottsTransitions2014, canova_competing_2016,nui_correlation_2018, roy2020quantum, song2021hybrid}. In this section, we give analytical arguments that reveal the subtleties which arise when addressing this question.

Vortex excitations in the system are extremely important, as their binding (unbinding) is related to the KT ordered (disordered) phases. Let us now discuss their role, and in particular the impact of nematic-hexatic phase locking at $T < T_\lambda$ as described by Eq.~\eqref{eq:2.6-locking_constraint} on their formation. For simplicity, we focus on the isotropic point $\Delta = 1$, but similar arguments can be given for other values of $\Delta$. It will be advantageous to refer to Hamiltonian~\ref{eq:1.2-model_xy}, which lends itself to a clean interpretation of the vortex defects. We provide a lexicon for the vortex excitations for the Hamiltonian of Eq.~\ref{eq:2.1-model_phi}, which is used extensively for technical reasons in the following sections. The same arguments can be made for the model of Eq.~\ref{eq:1.1-model_varphi} in the original variables, for which the minimal winding around a nematic (hexatic) vortex is given by $\pi$ ($2 \pi/6$), hence their respective names. For simplicity, we use $\lambda$ whenever we refer to the hexatic-nematic coupling term, irrespective of the particular model. 

In the absence of the coupling term ($\lambda=0$), hexatic and nematic systems undergo independent KT transitions (see Fig.~\ref{fig:figure_uncoupled_PD}). In the language of Hamiltonian~\ref{eq:1.2-model_xy}, the minimal charge of a hexatic (nematic) vortex is $q_\vartheta = 1/3$ ($q_\varphi = 1$) corresponding to a phase winding of $2 \pi q_{\vartheta}$ and $2 \pi q_\varphi$ around a vortex, respectively. Thus, at the KT transition, point vortices of charge $q_\varphi = \pm 1$ ($q_\varphi= \pm 1/3$) unbind in the nematic (hexatic) system. Free vortices limit the correlation lengths $\xi_6$ ($\xi_2$) for  $\vartheta$ ($\varphi$) to a finite value for all $T > T_{\text{KT}}$. Vortices are bound into pairs of total charge zero in the critical phase for $T < T_{\text{KT}}$, where the correlation lengths are infinite.

The situation is notably different at nonzero $\lambda$ and temperatures $T < T_\lambda$ below the locking crossover. While the domain of the angles reads $\vartheta \in [0, \frac{2\pi}{3})$ and $\varphi \in [0, 2\pi)$, the locking condition imposes a distinct constraint on the phase winding around nematic and hexatic vortices, such that

\begin{equation}\label{qlambda2} 
q_\vartheta \equiv q_\varphi + \frac{\Delta n}{3}  \; (\text{mod} \; 1) \quad\quad\quad (\lambda \neq 0) \;,
\end{equation}

\noindent where $\Delta n$ counts the changes in the Potts index $n$ as one loops around a vortex core. Note that the modulo operation originates from the definition of the Potts variables $n$, such that $n = -2 \equiv 1$. 

If the system is Potts disordered, one expects a sample to be swarmed by a network of domain walls of the $\sigma = \frac{2\pi}{3} n$ variable, as domains get smaller and smaller for temperatures above a Potts ordering temperature $T_3$. For such a system with global Potts disorder, $\Delta n \neq 0$ generically. Hence, Eq.~\ref{qlambda2} leads to $\frac13 \equiv  0 + \frac{1}{3} $, i.e. domain walls where $\Delta n = 1$ are necessarily attached to hexatic vortices. These domain walls are energetically costly due to the significant nematic gradient energy that arises across them. Proliferation of these vortices and their eventual unbinding leads to an hexatic  KT transition above the $T_3$ transition. This is the mechanism at play at small $J_2/J_6$ (or, alternatively, for small $\Delta$ in Eq.~\ref{eq:2.1-model_phi}), as it can be see in Fig.~\ref{fig:figure_intro2} (c), on the left side. 

In the presence of Potts order, one has that globally, $\Delta n = 0$. The solution to Eq.~\ref{qlambda2} is then $1 \equiv 1$ or $3(\frac13) \equiv 1$. Both cases correspond to nematic vortices, but there is a subtle difference between the two. In the first case, an hexatic vortex of charge \text{three} times its elementary charge is at the same site as a nematic vortex. This is a nematic point vortex.

The alternative is for the nematic vortex to correspond to a triad of hexatic vortices. Since each hexatic vortex is attached to a Potts domain wall, one simple way to solve such a triad is to link all three hexatic vortices via their domain wall, at it can be seen in Fig.~\ref{fig:figure_intro2} (e). In essence, the domain walls act as a binding force for the fractional vortices, i.e. the hexatic vortices, as they imply locally fractional nematic vortices.  Such a split vortex can be much more extended than its point vortex counterpart. In subsection~\ref{ssub:vortex_fractionalization_in_potts_ordered_phase}, we provide an analytical argument that extended nematic vortices are lower energy than their point vortex. Even though the domain walls can be quite energetic, vortices of charge triple that of their elementary charge are extremely costly due to the phase winding around the vortex core.

Note that if one uses the convention of Eq.~\eqref{eq:2.1-model_phi}, where both $\theta$ and $\phi$ are within the interval $[0, 2 \pi)$, the minimal phase winding around a vortex is given by $2 \pi$ in both cases, corresponding to charge $q_\theta = 1$ and $q_\phi =1$ elementary vortices. The relative locking between hexatic and nematic angles is described then by Eq.~\eqref{eq:2.6-locking_constraint}, leading to the constraint that 

\begin{equation}\label{qlambda3} 
\frac{q_\theta}{3} \equiv q_\phi + \frac{\Delta n}{3}  \; (\text{mod} \; 1) \quad\quad\quad (\lambda \neq 0) \;.
\end{equation}

\noindent Thus, a nematic vortex with minimal $2 \pi$ phase winding must be accompanied by a hexatic vortex of charge $q_\theta = 3$ that exhibits a phase winding of $2 \pi q_\theta = 6 \pi$. In both descriptions, a nematic vortex of minimal charge $q_\phi$ necessarily pairs with a hexatic vortex whose charge is three times larger than its minimal charge. The hexatic vortex is then a situation where $q_\theta = 1$ and $\Delta n = 1$, which means that there is a Potts domain wall due to the $2\pi/3$ mismatch in the $\phi$ variable. Similarly, the composite nematic vortex has $q_\phi = 1$ and three $q_\theta = 1$ vortices, such that $(1 + 1 + 1)/3 \equiv 1$, with each hexatic vortex confined via the attractive domain wall. For all the following sections, we use the model presented at Eq.~\ref{eq:2.1-model_phi}. Translation between the different vortex formalism (Eq. \ref{eq:1.2-model_xy} vs \ref{eq:2.1-model_phi}) can be done through this subsection. 

In other words, in this model, the transition to the relative Potts phase corresponds to confinement transition of fractionalized $q_\varphi = 1/3$ nematic vortices, which drives the formation of a bound state of the three attached elementary $q_\vartheta = 1/3$ hexatic vortices through the constraint of Eq.~\ref{qlambda2}. In the absence of free elementary hexatic vortices, a Potts order parameter can be well defined even in the presence of nematic vortices. In contrast, no independent $\mathbb{Z}_3$ degree of freedom exists below the nematic KT transition, as the potential $\lambda$ acts as a uniaxial field for the hexatic degrees of freedom. Furthermore, elementary hexatic vortices are attached to ``dangling'' domain walls which, if they were not confined, would fully destroy the LR Potts order that is set in the system. The composite vortices of total elementary nematic charge are then the only ones allowed in the relative $\mathbb{Z}_3$ ordered state. We cover the energetics of the extended nematic vortices in the following subsection.

This transition is expected to lie in the 2D Potts universality class, which is characterized by the critical exponents $\alpha = 1/3, \beta = 1/9, \gamma = 13/9, \nu = 5/6$~\cite{wu1982potts}. Note in particular that the specific heat exponent $\alpha$, such that $c \propto t^{-\alpha}$, leads to a pronounced divergence of the specific heat distinct from the smooth behavior observed for KT transitions with a broad hump at about $1.1  T_{\text{KT}}$~\cite{chaikin1995principles}. We will exploit this notable difference to distinguish between Potts and KT transitions in our Monte-Carlo simulations in section~\ref{sec:Monte_Carlo_results}.

\subsection{Vortex fractionalization and the extended vortices} 
\label{ssub:vortex_fractionalization_in_potts_ordered_phase}

Extended nematic $q_\varphi$ vortices are the free ``elementary excitations'' in the system at temperatures below $T_\lambda$. Lower charge vortices are held together by domain wall strings with finite tension, arising from the gradient energy cost of the domain wall due to the ``fractional'' hexatic vortex they contain.
This observation immediately questions the possibility that a KT transition can take place below the Potts transition, because \emph{point-like} vortices are expected to bind at a temperature above a possible Potts transition $T_3$. Specifically, setting $\nabla \vartheta = \nabla \varphi$ once below $T_\lambda$ (alternatively, setting $\nabla \theta = 9 \nabla \phi$ due to Eq.~\eqref{eq:2.6-locking_constraint}) leads to $T_{\text{KT}, \rm nem}^{\lambda \neq 0} = 10 \, T_{\text{KT}}^{\lambda = 0} \approx \frac{10 \pi}{2} J$. The estimated KT transition temperature is thus even larger than $T_{\lambda} \approx \frac{8 \pi}{9}J$ (which bounds $T_3$ in the small $\lambda$ regime).

This estimate, however, leaves out the possibility of vortex fractionalization and the emergence of extended vortices \cite{nelson1980solid,dierker1986consequences}. Indeed, it is well known that higher charge vortices can reduce their (gradient) energy by splitting into multiple lower charge objects. In addition, hexatic and nematic degrees of freedom are continuous, therefore the domain walls separating regions with different Potts variable $n_i$ can acquire a finite width $\xi_{dw}$. The width $\xi_{dw}$ is determined by a balance between the cost of violating the locking constraint in Eq.~\eqref{eq:2.6-locking_constraint} imposed by the $\lambda$ term and the gain in gradient energy by distributing the angle mismatch over a finite length. The width of the domain walls approaches the minimal size of the lattice constant only as the renormalized $\lambda \gg J$. For example, splitting a $q_\theta = 3$ vortex into three $q_\theta = 1$ vortices reduces the gradient energy by a factor of $3^2 - 3 = 6$~\cite{radzihovsky2008superfluidity} (similarly, splitting a $q_{\vartheta} = 1 = 3(\frac13)$ vortex).
Previous mean-field studies of continuous models with the same symmetry have shown that such an arrangement is energetically beneficial~\cite{babaevphase2004,nitta2012baryonic, kobayashi2020z, kobayashi2020vortex, kobayashi2019berezinskii}. 

Here, this splitting, however, implies fractionalization of the joint nematic vortex into three vortices of fractional charge $q_\phi = 1/3$. These are held together by domain wall strings where $\Delta n = \pm 1$. The competition of gradient and domain wall energy results in an extended vortex core~\cite{radzihovsky2008superfluidity}. By comparing the energy of a point vortex ($3 = 3(1)$) to a split vortex ($1 + 1 + 1 = 3(1)$, which looks like a point vortex at distances larger than the core), we arrive at the conclusion that it is energetically favored to split the vortex. For a circular geometry as in Fig.~\ref{fig:figure_intro2} (e), this leads to an optimal radius of $ R \simeq 2.5 a$ for the value of the coupling $\lambda$ used in this work, or a split vortex with an area $6.25$ times greater than its point vortex counterpart. Thorough derivation of this result is presented in Appendix~\ref{appendix2}. Importantly, the unbinding of extended vortices is known to occur at a lower temperature than the unbinding of point-like vortices~\cite{fellowsUnbindingGiantVortices2012}. This can be understood from the fact that the initial value of the vortex fugacity increases by a factor of $(R/a)^2$, where $a$ is the microscopic lattic scale. For $R \gg a$, the KT transition temperature scales as $T_{\text{KT}} \sim 1/\ln[(R/a)^2]$ and is thus significantly reduced
in the case of large vortex core sizes. The crucial open question is whether it is reduced to a value below the Potts transition temperature. To address this question in an unbiased way, we have performed large scale classical Monte-Carlo simulations, which will be discussed next. 


\subsection{Previous numerical studies}
\label{ssec:previous_studies}


The particular model presented in Eq.~\ref{eq:2.1-model_phi} was first studied in two dimensions using Monte-Carlo techniques \cite{jiangMonteCarloSimulation1993, jiang1996monte}, yet with a focus on a different region of the phase diagram ($\Delta \ll 1$) and at substantially smaller system sizes than presented here. A three-dimensional version of the model was studied as well~\cite{ghanbari2005monte,shahbazi2006emergence}. A related model, with $\text{O(2)} \times \mathbb{Z}_2$ symmetry instead of our $\text{O(2)} \times \mathbb{Z}_3$ symmetry, associated with atom-molecular mixing, was studied extensively in two dimensions~\cite{de2016multiple, de2017quantum}. The $\Delta \simeq 1$ regime was however not the focus of those studies. 

The community has been more focused on the infinite coupled limit. Such generalized XY models can be written as 

\begin{align}
    \mathcal{H}_{\infty} &=  -  \sum_{\av{i,j}} \big[\Delta\cos (\delta \phi_{ij}) + (2 - \Delta)\cos (p\delta \phi_{ij}) \big] \,, \label{eq:model-lambda-inf}
\end{align}

\noindent where $\delta \phi_{ij} = \phi_i - \phi_j$, with $p=3$ corresponding to the $\lambda \rightarrow \infty$ limit of Eqs.~\ref{eq:1.2-model_xy} and \ref{eq:2.1-model_phi}, and $p=2$ is obtained through the same limit for the atomic-molecular mixing model. These two limiting model were studied via renormalization group studies \cite{lee1985strings,korshunov1985possible,korshunov86,granato1986critical,shiBosonPairingUnusual2011, serna2017deconfinement}, Monte-Carlo techniques \cite{CarpenterChalker89,poderosoNewOrderedPhases2011, canovaKosterlitzthoulessPottsTransitions2014, canova_competing_2016,nui_correlation_2018,vzukovivc2018multiple}, matrix-product states \cite{song2021hybrid}, bosonization \cite{bonneshalfvortex2012,roy2020quantum}. They all share a common phase diagram structure \cite{korshunov2006phase}, with a well understood regime at $\Delta \ll 1$ with a $p$-state discrete transition at $T_d$ (Ising or Potts) below a Kosterlitz-Thouless transition at $T_{KT}$ where fractional defects unbind. For $\Delta \simeq 1$, the available evidence supports a single transition temperature corresponding to both confinement/deconfinement transition of fractional vortices and an unbinding of integer charge vortices \cite{shiBosonPairingUnusual2011, serna2017deconfinement, roy2020quantum}. From our analysis of the composite vortex in section~\ref{ssub:vortex_fractionalization_in_potts_ordered_phase}, we find that there is no gain at $\lambda \rightarrow \infty$ for an extended vortex over a point one. This would likely collapse the two transitions we expect into one in that limit, such that they would be unseparable with current numerical accuracy. This is partly why our numerical study is focused on the coupled model in the intermediate regime, so as to see the postulated effect of the formation of composite vortices.

We note that in the case of the fully frustrated XY model (FFXY), which represents a periodic array of Josephson junctions with half a quantum flux through it, one rather has two $\text{O(2)}$ order parameters coupled via a term $\sim \cos{[2(\theta_i -\phi_i)]}$. For this model, extensive analytical \cite{halseytopological1985,choi1985phase, yosefin1985phase, granato1986critical,granato1991phase} and Monte-Carlo simulations \cite{teitel1983phase, lee1991monte, olsson1995two, jeon1997double, simon19972d,loison2000monte, olsson2005kink, hasenbusch2005multicritical} were able to distinguish a clear regime where the relative $\mathbb{Z}_2$ symmetry breaking happens at a higher temperature than the global KT transition. It was also shown that the presence of the Ising line above the KT transition leads to the prospect of an emergent supersymmetry at the Ising-KT multicritical point \cite{huijseemergent2015}. This success was due to the power of modern implementation of Monte-Carlo algorithms, as well as the ever-increasing computing power available, further motivating our reexamination of this classic hexatic-nematic problem.

\section{Monte-Carlo simulation: algorithm and observables}
\label{sec:Monte_Carlo_algorithm}
\begin{figure*}
    \centering
    \includegraphics[width=\linewidth]{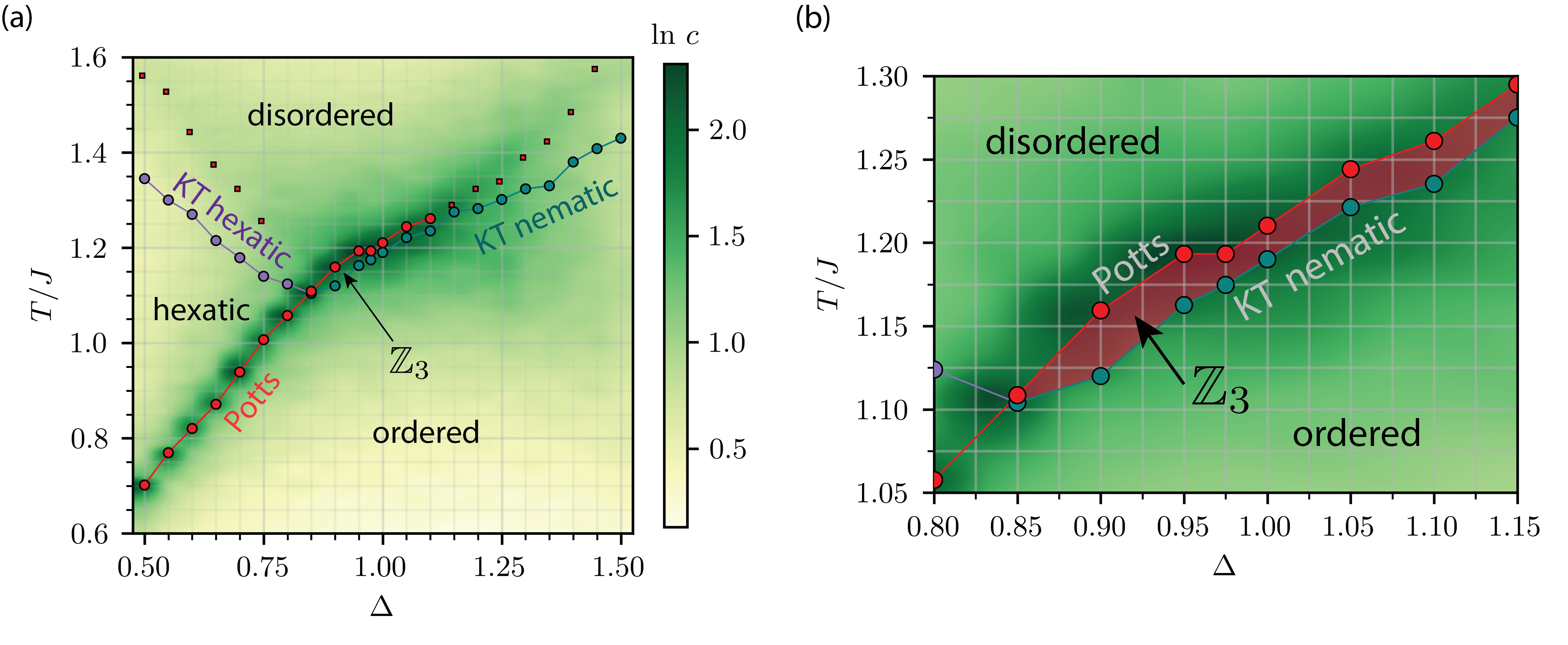}
    \caption{(a) Numerical phase diagram of coupled hexatic-nematic XY model in Eq.~\eqref{eq:2.1-model_phi} as a function of temperature $T$ and ratio of exchange couplings $\Delta = J_2/J$, obtained from classical Monte-Carlo simulations. The hexatic-nematic coupling strength is set to $\lambda = 2.1 J$ and $J = \frac12 (J_2 + J_6)$. Background color density denotes specific heat per site $c$ at system size $L=40$ (see logarithmic color bar).
    We distinguish four different phases: (i) a disordered phase with purely short-range correlations, (ii) a hexatic phase with algebraic correlations of $\langle \cos(\theta_i - \theta_j) \rangle \sim |i-j|^{-\eta_6}$, (iii) a $\mathbb{Z}_3$ Potts phase with long-range correlations $\langle \cos(\sigma_i - \sigma_j) \rangle \rightarrow \text{constant} > 0$ as $|i-j|\rightarrow \infty$ with $\sigma_i = (\theta_i - 3 \phi_i)/3$, but short-range correlations of hexatic and nematic angles, and (iv) an ordered phase with algebraic correlations of both hexatic and nematic degrees of freedom. Here, $\langle \mathcal{O} \rangle$ denotes the thermal average.
    The KT transition temperatures (purple and blue dots) are obtained from the total stiffness jump criterion. The size of the jump is used to distinguish between a nematic KT transition, where $\rho(T_{\text{KT}}) = 2 T_{\text{KT}}/\pi$ (unbinding of $q_\phi = 1$ vortices), and a hexatic KT transition, where $\rho(T_{\text{KT}}) = 18 T_{\text{KT}}/(\pi)$] (unbinding of $q_\theta = 1$ vortices). Note that we extrapolate to infinite system size (see Fig.~\ref{fig:figure_Delta_1}).
    The Potts transition temperature (red dots) is determined from the maximum of the specific heat $c$ at $L=40$, except at $\Delta = 0.95, 1.0$, where $T_3$ is obtained by fitting $c$ to the Potts scaling form $c = t^{-\alpha}$ with $t = |T- T_3|/T_3$ and $\alpha = 1/3$ (see Figs.~\ref{fig:figure_small_Delta} and~\ref{fig:figure_Delta_1}). Consistent Potts transition temperatures are obtained from a universal scaling analysis of the magnetization, susceptibility and specific heat, as shown in Fig.~\ref{fig:figure_scaling_analysis}. The red squares for $\Delta \geq 1.5$ track a broad maximum of $c$ associated with the KT nematic transition at $T_2$, demonstrating that the nature of the upper phase transition changes from Potts to KT at $\Delta = 1.15$. On the left side of the phase diagram, the red squares correspond to the second, lower maxima of the specific heat corresponding the proliferation of hexatic vortices above the hexatic KT transition. (b) Zoom into region with relative $\mathbb{Z}_3$ Potts order. The Potts phase exists for $0.9 \leq \Delta \leq 1.15$ and extends over a region $\Delta T/J \approx 0.005$.
    }
    \label{fig:figure_coupled_PD}
\end{figure*}

In this section, we describe the algorithm used for our large-scale parallel tempering Monte-Carlo (MC) simulations of the model in Eq.~\eqref{eq:2.1-model_phi}. We give details in subsection~\ref{sub:technical_details_of_monte_carlo_algorithm}, and introduce the various observables that are measured to obtain the phase diagram of the model, presented in Fig.~\ref{fig:figure_coupled_PD}, in subsection~\ref{ssec:observables_phases_transitions}. The detailed investigation of the full phase diagram is presented in Sec.~\ref{sec:Monte_Carlo_results}. All data and code needed to generate the results in this paper are available online~\cite{githubref}.

\subsection{Technical details of Monte-Carlo algorithm} 
\label{sub:technical_details_of_monte_carlo_algorithm}
The employed MC algorithm combines a standard parallel tempering update \cite{marinari_simulated_1992, hukushima1996exchange} with a generalized Wolff cluster algorithm adapted to coupled XY models~\cite{wolff_collective_1989, de2016multiple}. The simulations were performed on a square lattice with periodic boundary conditions of $N = L \times L$ sites, with $L$ ranging from $8$ to $420$. For system sizes $L \leq 200$ ($L > 200$), we simulate 40 (64) configurations in parallel at different, geometrically spaced temperatures between $T_{\text{min}} = 0.5J$ and $T_{\text{max}} = 1.6 J$ to obtain a general overview of the phase diagram. When looking directly at the nature of a phase transition at $T_c$, we always used a new temperature range such that $(T_{\text{min}},T_{\text{max}}) = (0.95\: T_c, 1.05 \: T_c)$. The generalized Wolff algorithm takes into account that due to the coupling term, $\lambda \sum_i \cos(\theta_i - 3\phi_i)$, one must flip the hexatic and nematic angles, $\theta_i$ and $\phi_i$, in an anisotropic manner in order to explore both global O(2) and $\mathbb{Z}_3$ symmetries. The Wolff clusters are therefore built as follows: one first starts by randomly selecting a flip direction $\eta \in [0, 6\pi)$, a site $i$ and the type of variable, $\theta_i$ or $\phi_i$.
Depending on which variable was chosen, we then apply the flip rule
\begin{subequations}
\begin{align}
\theta_i & \rightarrow \theta'_i = - \theta_i + \eta \\
\phi_i & \rightarrow \phi'_i =  - \phi_i + \frac{\eta}{3} \,.
\end{align}
\end{subequations}
Subsequently, one grows the cluster by investing all five bonds connected to the chosen variable at site $i$. These connect to the four nearest-neighbors of the same variable and to the other variable at the same site $i$. For each of these bonds, one calculates the energy cost $\Delta E$ of flipping the other member of the bond as well. This member is flipped and the site added to the cluster with probability $p = 1- \min\{1, \exp{(-\beta \Delta E)}\}$, where $\beta = 1/T$. One then proceeds in the same way for all new members of the cluster until all outgoing bonds of the cluster have been considered. Note that the clusters invades both $\{\theta_i\}$ and $\{\phi_i\}$ variables.

We perform the simulations for at least $4 \times 10^5$ MC steps, where each step consists of a parallel-tempering and a generalized Wolff move. To ensure thermalization, we discard the first half of obtained configurations and measure thermodynamic observables only during the second half of MC steps. We introduce the different observables that we measure next. Finally, error bars are obtained using the standard jackknife method~\cite{efron1982jackknife}.

\subsection{Observables, phases and phase transitions}
\label{ssec:observables_phases_transitions}
To distinguish the different phases of the model, we investigate several thermodynamic observables that can be grouped into three classes. First, we measure observables related to the energy fluctuations in the system, the specific heat per site $c$ and the Binder cumulant of the energy $B_E$:
\begin{subequations}
\begin{align}
    c &= \frac{\langle \mathcal{H}^2 \rangle - \langle \mathcal{H} \rangle^2}{N T^2 J^2}
    \label{eq:3.1a}
    \\
    B_E &=  \frac{\langle \mathcal{H}^4\rangle}{\langle \mathcal{H}^2 \rangle^2} - 1  \,,
    \label{eq:3.1b}
\end{align}
\label{eqs:3.1}
\end{subequations}
where $\mathcal{H}/J$ is defined in Eq.~\eqref{eq:2.1-model_phi}. A characteristic signature of second-order phase transitions are strong fluctuations of the energy close to the critical point. The specific heat diverges as $c \propto t^{-\alpha}$ with $t = |T-T_c|/T_c$, and $B_E$ exhibits a sharp local maximum at $T_c$~\cite{Binder1981, martinos2005finite, velonakis2014critical, velonakis_efficient_2015}. In contrast, energy fluctuations are much less pronounced and more broadly distributed at a KT transition. Specifically, at a temperature of about $T = 1.1\, T_{KT}$, the specific heat shows a rounded bump and $B_E$ exhibits a change in slope, which signals the thermal generation of vortex excitations. Both observables $c$ and $B_E$ are, therefore, well suited to distinguish between the second-order Potts phase transition (of $n_i$) and the KT transitions of hexatic and nematic variables $\theta_i$ and $\phi_i$. In particular, the sustained presence of a sharp local maximum in $B_E$ for all system sizes is a clear signature of a second-order phase transition.

The second class of observables we study are magnetizations and their susceptibilities that characterize the different phases
\begin{subequations}
\begin{align}
    m_\theta = \braket{M_{\theta}} &= \frac{1}{N} \sum_i \braket{\cos\theta_i}
    \label{eq:3.2a} \\
   m_\phi = \braket{M_{\phi}} &= \frac{1}{N} \sum_i \braket{\cos \phi_i}
    \label{eq:3.2b} \\
    m_\sigma = \braket{M_{\sigma}} &= \frac{1}{N} \sum_i \braket{\cos\sigma_i}\,.
    \label{eq:3.2c}
\end{align}
\label{eqs:3.2}
\end{subequations}
Here, we have introduced the $\mathbb{Z}_3$ Potts variable as the following:
\begin{subequations}
\begin{align}
   \sigma_i &= \frac{2 \pi}{3} n_i = \frac13(\theta_i - 3\phi_i)
\end{align}
\label{eq:def_sigma}
\end{subequations}
\noindent such that the hexatic-nematic coupling Hamiltonian term reads $\lambda \cos[3\sigma_i]$.
We also measure the associated susceptibilities and Binder cumulants ($a = \theta, \phi, \sigma$):
\begin{subequations}
\begin{align}
    \chi_a &= \frac{N}{T} \Bigl(\langle M_a \rangle^2 - \langle M_a^2 \rangle\Bigr)
    \label{eq:3.3a} \\
    B_a &=  1 - \frac{\langle M_a^4\rangle}{3\langle M_a^2 \rangle^2}
    \label{eq:3.3b}
\end{align}
\label{eqs:3.3}
\end{subequations}
Below we perform a scaling analysis of $m_a$, $\chi_a$ and $c$ in order to extract the critical exponents of the observed second order phase transition. The Binder cumulants of the magnetizations undergo step-like transitions from a value of $1/3$ at high temperature to a value of $2/3$ at low temperature at both KT and second-order phase transitions.

Finally, the third class of observables that we investigate are spin stiffnesses $\rho_\alpha$, which describe the free energy change of the system under a uniform phase twist (along a given lattice direction $\alpha$). When hexatic and nematic variables are coupled to each other by nonzero $\lambda$, one cannot separate contributions arising from each individual variable and thus only extract the total stiffness of the system. In order to fulfill the potential term constraint~\eqref{eq:2.6-locking_constraint}, we apply a uniform phase twist of the form
\begin{align}
\bigl(\phi_{i + \hat{\alpha},i} , \theta_{i + \hat{\alpha},i}\bigr) \rightarrow \bigl(\phi_{i + \hat{\alpha},i} + \psi, \theta_{i + \hat{\alpha},i} + 3\psi \bigr) \,,
\label{eq:3.4}
\end{align}
where $\phi_{ji} = \phi_j - \phi_i$, $\theta_{ji} = \theta_j - \theta_i$,
Here, $\psi$ is uniform across the system and $\hat{\alpha}$ corresponds to either $\hat{x}$ or $\hat{y}$ lattice direction. Note that the phase twist that is applied to the hexatic angle $\theta_i$ across each bond is three times larger than the twist applied to the nematic angle.

The spin stiffness for a twist along direction $\alpha$ is defined as $\rho^{(\alpha)} = \frac{1}{N} \frac{\partial^2 F}{\partial \psi^2}$ and following a standard derivation~\cite{sandvikComputationalStudiesQuantum2010}, one arrives at the explicit expressions
\begin{subequations}
\begin{align}
\rho &= \frac12 \bigl(\rho^{(x)} + \rho^{(y)}\bigr) \\
\rho^{(\alpha)} &=\frac{1}{N}  \av{ \mathcal{H}^{(\alpha)} }  - \frac{\beta}{N} \left[ \av{(I^{(\alpha)})^2}- \av{I^{(\alpha)}}^2 \right] \\
\mathcal{H}^{(\alpha)} &= \Delta \sum_{\langle i,j \rangle_\alpha}  \cos{\phi_{ij}} + 9 (2-\Delta) \sum_{\langle i,j \rangle_\alpha}   \cos{ \theta_{ij}}  \\
I^{(\alpha)} &= \Delta \sum_{\langle i,j \rangle_\alpha}   \sin{\phi_{ij}} + 3 (2-\Delta) \sum_{\langle i,j \rangle_\alpha}   \sin{ \theta_{ij} }\,.
\end{align}
\label{eqs:stiffness}
\end{subequations}
Here, we have defined the total stiffness $\rho$ that is averaged over both lattice directions. The summation $\av{i,j}_\alpha$ runs over all bonds along direction $\alpha$. For $\lambda=0$, cross correlations between hexatic and nematic variables are absent and the total stiffness decomposes into the sum $\rho(\lambda=0) = \rho_2 \Delta  + 9 \rho_6(2-\Delta) $, where $\rho_2$ ($\rho_6$) is the stiffness of the nematic (hexatic) system for a twist angle of $\psi$. Note that the factor of $9$ in the $\rho_6$ part is due to the fact that we apply a uniform twist angle of $3 \psi$ to the hexatic angles.

It is well known that the discontinuous jump of the spin stiffness at the KT transition can be directly associated with the charge $q$ of the unbinding vortices (phase winding of $2 \pi q$ around the vortex)~\cite{weberMonteCarloDetermination1988,hubscherStiffnessJumpGeneralized2013,canovaKosterlitzthoulessPottsTransitions2014}. In particular, one finds
\begin{equation}
    \rho\bigl[T_{\text{KT},q}^{-}\bigr] = \frac{2T_{\text{KT},q}^{-}}{\pi q^2}
    \label{eq:3.7-KT_jump}
\end{equation}
just below the transition, while the stiffness vanishes above the transition. The hexatic-nematic model of Eq.~\eqref{eq:2.1-model_phi} supports two types of vortex excitations, associated with the hexatic and nematic angles. Because of the choice we made in Eq.~\eqref{eq:3.4} to apply the phase twist uniformly on the nematic $\phi$ variables, then the type of vortex unbinding in $\phi$ dictates the value of the stiffness jump. A KT transition dominated by the unbinding of $q_{\phi} = \pm 1$ vortices 
results in a normal jump of the total stiffness $\rho$ by $2T_{\text{KT},2}/\pi$. We thus denote a KT transition characterized by this jump value as ``nematic KT'', $T_{\text{KT},2} \equiv T_{2}$. On the other hand, a KT transition driven by the unbinding of integer charge $q_{\theta} = \pm 1$ hexatic vortices, accompanied by their corresponding Potts domain walls, results in a jump of the total stiffness by $18 T_{\text{KT}, 6}/\pi$. A KT transition characterized by this larger jump value is denoted as ``hexatic KT'', $T_{\text{KT},6} \equiv T_{6}$. To summarize, the location of stiffness jumps is used to determine the KT transition temperature $T_{\text{KT}}$, and the height of the jump provides clear evidence on the type of vortex unbinding that occurs across the transition.

In the following, we will use these observables to identify the different phases and phase transition universality classes presented in Fig.~\ref{fig:figure_coupled_PD}.

\section{Computational results}
\label{sec:Monte_Carlo_results}
Here we present results of extensive MC simulations of the model~\eqref{eq:2.1-model_phi}. Our main result is the phase diagram shown in Fig.~\ref{fig:figure_coupled_PD} as a function of temperature $T$ and coupling constant ratio $\Delta= J_2/J$ with $J = \frac12 (J_2 + J_6)$. The phase diagram is obtained for fixed hexatic-nematic coupling strength $\lambda = 2.1$. Crucially, it shows a region around to $\Delta = 1$, where the Potts phase transition $T_3$ lies above the nematic KT transition $T_{\text{KT, nem}}$, which is highlighted in Fig.~\ref{fig:figure_coupled_PD}(b). Although the two transitions occur close to each other and the ratio of transition temperatures reads $T_3/T_{\text{KT}}(\Delta = 1) = 1.007$, a detailed numerical analysis presented below shows that $T_3 > T_{\text{KT}}$ with a high degree of statistical confidence.
This numerically demonstrates that LR order in the emergent Potts $\mathbb{Z}_3$ variable $n_i$, as defined related to $\sigma_i$ in Eq.~\ref{eq:def_sigma},
exists even in a region with finite-range correlations of the hexatic and nematic degrees of freedom. This suggests that the HO phase that is experimentally observed in 54COOBC films is characterized by LR Potts order and the sharp specific heat divergence is associated with a 2D Potts phase transition. In the following subsections, we separately discuss the different regions in the phase diagram in the order of increasing values of $\Delta$.
\begin{figure}[tb]
    \centering
    \includegraphics[width=\linewidth]{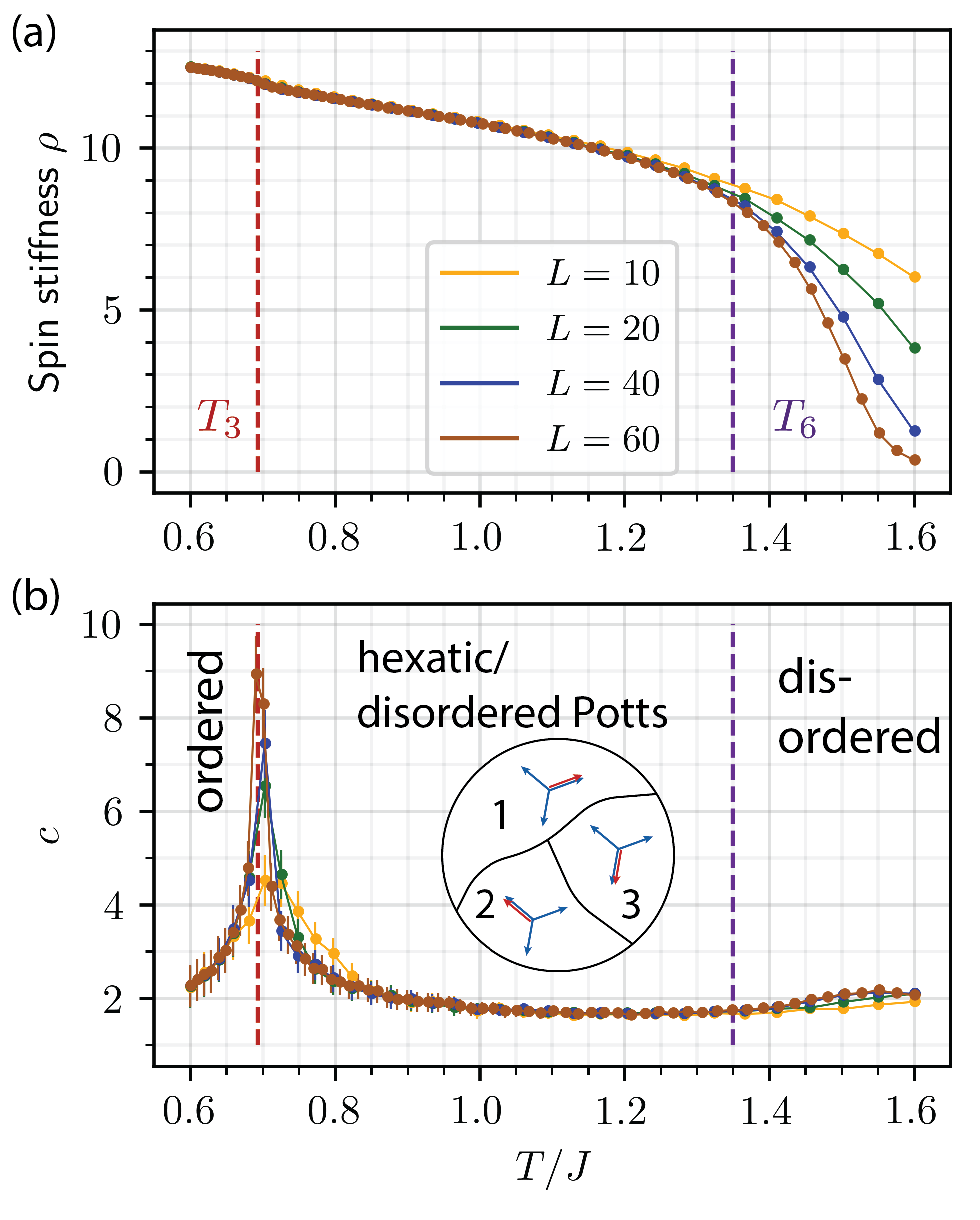}
    \caption{
    (a) Total spin stiffness $\rho$ defined in Eq.~\eqref{eqs:stiffness} as a function of temperature $T/J$. Upturn at $T_6$ signals KT transition into the hexatic phase. The transition temperature is determined most accurately using the universal jump criterion $\rho(T_6^-) = 18 T_6/\pi$, which yields $T_6(\Delta = 0.5) = 1.35 \pm 0.01$.
    (b) Specific heat per site $c$ as a function of temperature $T/J$ for $\Delta = 0.5$. While $c$ exhibits a broad maximum above the hexatic KT transition at $1.1 \, T_6$, it develops a sharp peak at the lower transition at $T_3$, which increases with system size $L$. This is characteristic of Potts phase transition in 2D, where $c \sim t^{-\alpha}$ with exponent $\alpha = 1/3$. We extract the transition temperature $T_3(\Delta = 0.5) = 0.693 \pm 0.005$ from the location of the maximum of $c$ for $L=60$.
    Inset depicts the three $\mathbb{Z}_3$ domains with different relative hexatic-nematic ordering in the hexatic phase. Long-range $\mathbb{Z}_3$ order only develops at the lower $T_3$ transition.
    }
    \label{fig:figure_small_Delta}
\end{figure}

\subsection{Upper KT and lower Potts transition at $\Delta <0.9$} 
\label{sub:upper_KT_small_Delta}
As shown in the phase diagram in Fig.~\ref{fig:figure_coupled_PD}(a), for $\Delta < 0.9$ and starting from the disordered high-temperature phase, the system first develops hexatic QLRO across a KT transition at temperature $T_6$. This transition corresponds to an unbinding of hexatic vortices with charge $q_\theta = \pm 1$. Since $T_6 < T_\lambda$, i.e. the temperature at which the $\lambda$ coupling becomes relevant, each hexatic vortex is found at the end of a domain wall string, and can also be thought of as a fractional nematic vortex of $q_\phi = 1/3$. As $T_6 > T_3$, Potts domain walls have proliferated already at lower temperatures. As shown in Fig.~\ref{fig:figure_small_Delta}(a) the unbinding of fractional nematic vortices results in a jump of the total stiffness equal to $\rho(T_6^-) = \frac{18}{\pi} T_6$. We determine $T_6$ by extracting $T_6(L)$ for different system sizes $L$ using this criterion and then extrapolating to infinite system size.
As shown in Fig.~\ref{fig:figure_small_Delta}(b), the specific heat $c$ exhibits only a broad maximum at about $1.1\, T_6$, as expected from a KT transition. In contrast, $c$ develops a pronounced peak at lower temperature $T_3$, which grows with system size $L$. Such a power law singularity is expected, for example, at a 2D Potts phase transition, where $c \sim t^{-\alpha}$ with $\alpha = 1/3$ and $t = (T-T_3)/T_3$.
\begin{figure*}
    \centering
    \includegraphics[width=\linewidth]{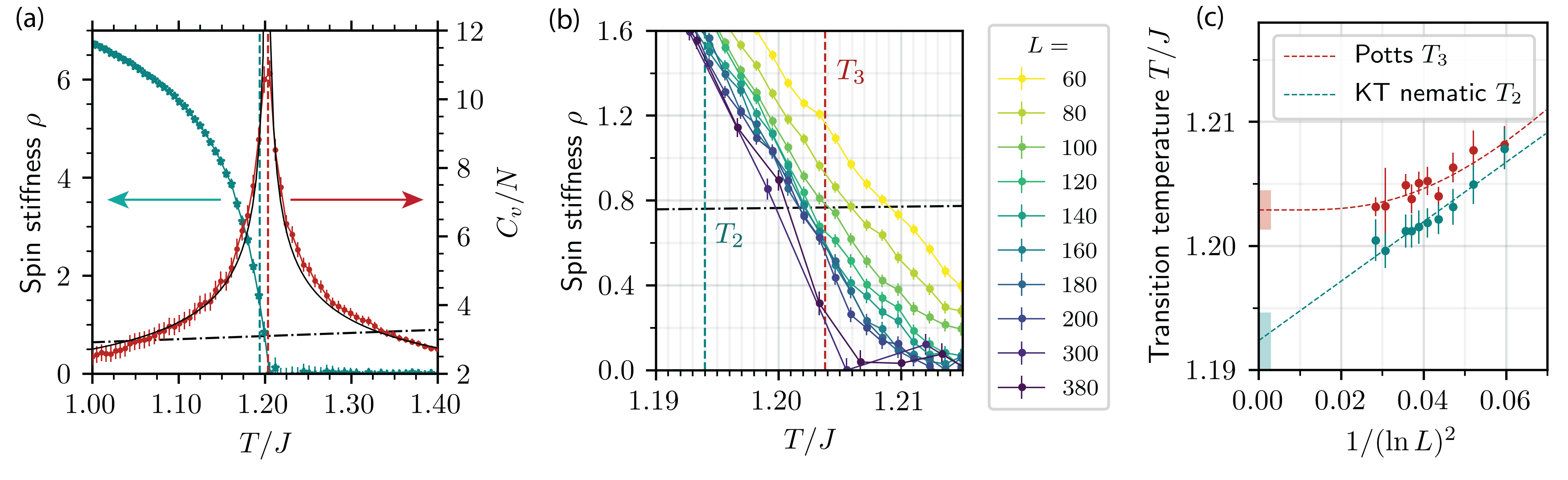}
    \caption{
    (a) Spin stiffness $\rho$ and specific heat per site $c$ as a function of temperature $T/J$ at $\Delta = 1.0$ and for linear system size $L = 300$. Dashed vertical lines denote the asymptotic transition temperature $T_2\equiv T_2(\infty)$ (blue) and $T_3 \equiv T_3(\infty)$ (red) from extrapolation to $L \rightarrow \infty$ (see panel (c)).The divergence of $c$ at $T_3$ occurs close to the upturn of $\rho$ at $T_2$. The solid black line is a fit of $c$ to $t^{-\alpha}$ with $\alpha = 1/3$, which allows the extraction a Potts transition temperature $T_3(L)$. The crossing of the dot-dashed black line $2T/\pi$ with $\rho$ allows determining the KT transition temperature $T_2(L)$. 
    (b) Spin stiffness as a function of temperature $T/J$ for $\Delta = 1.0$ and different system sizes between $L=60$ and $L=380$. The crossing of $\rho$ with the dot-dashed black line at $T_2(L)$ shifts to lower temperatures as the system size $L$ increases. Dashed vertical lines follow the same convention as in panel (a).
    (c) Evolution of Potts $T_3(L)$ and nematic KT $T_2(L)$ transition temperatures with system size $L$. Extrapolation to infinite system size is obtained by best fits to $T_2(L) = T_2(\infty) + a/(\ln L)^2$ and $T_3(\infty) = T_3(L) + a'L^{-1/\nu}$ with Potts exponent $\nu = 5/6$ (cyan and red dashed lines).
    We find $T_2(\infty) = 1.194 \pm 0.002$  and $T_3(\infty)=1.2022 \pm 0.0005$, and non-universal constants $a=0.21 \pm 0.05, a'= 1.00 \pm 0.16$. Shaded boxes denote best estimate for thermodynamic transition temperatures with a confidence of one standard deviation. This demonstrates that the Potts transition at $T_3$ occurs above the nematic KT transition at $T_2$.
    }
    \label{fig:figure_Delta_1}
\end{figure*}

\begin{figure}
    \centering
    \includegraphics[width=\linewidth]{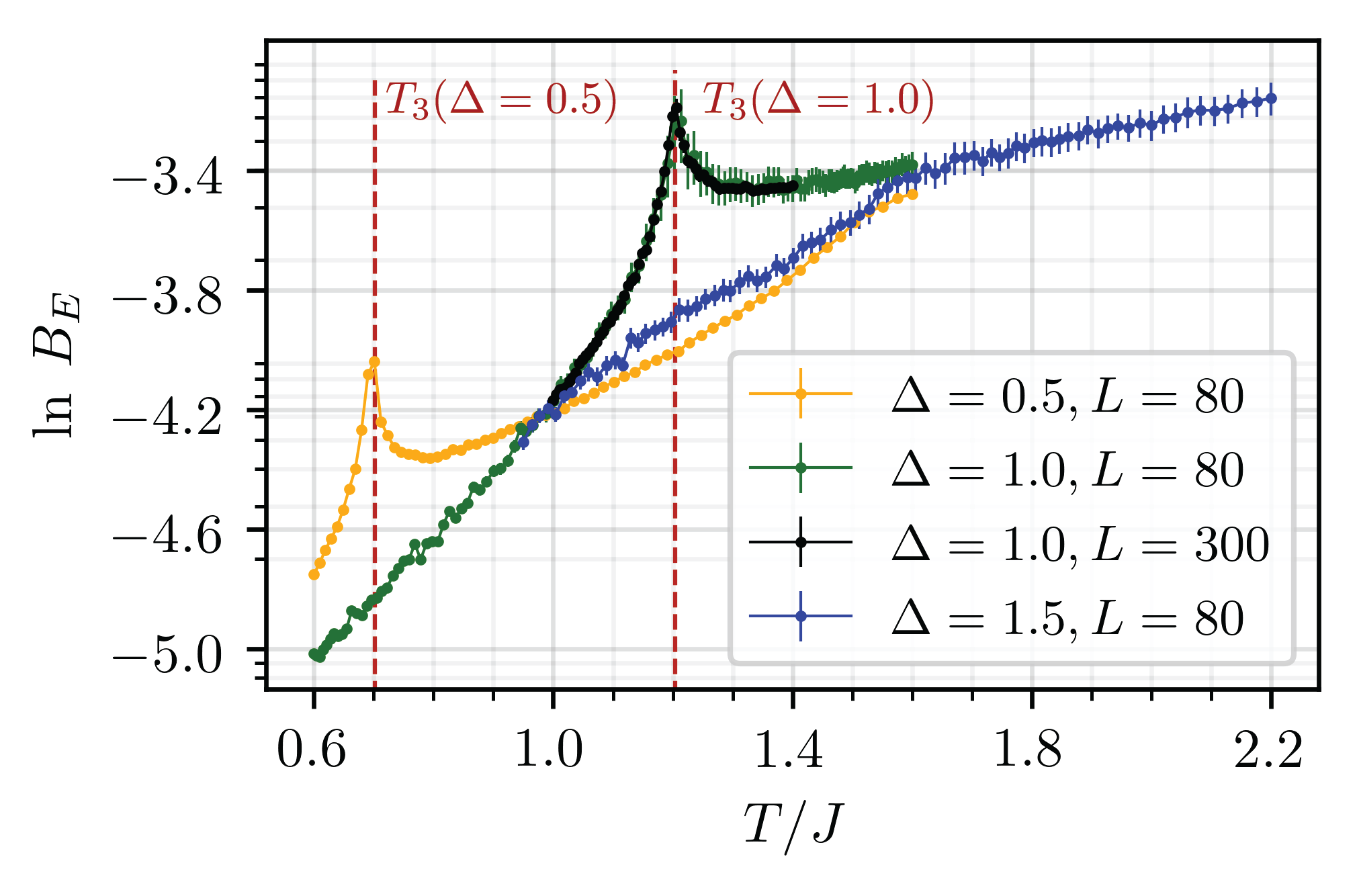}
    \caption{Binder cumulant of the energy $B_E = \frac{\langle E^4\rangle}{\langle E^2 \rangle^2} - 1$ as a function of temperature $T/J$ for different values of $\Delta$. The system size was set to $L = 80$.
    At $\Delta = 0.5$, the energy Binder develops a sharp peak at the Potts transition $T_3$ (vertical dashed line) as obtained in Fig.~\ref{fig:figure_coupled_PD}(b). The apparent change in slope at around $T = 1.5 J$ coincides with the characteristic broad hump in the specific heat at around $1.1 \, T_{6} = 1.5$ with $T_6 = 1.35 J$. At $\Delta = 1.0$, the sharp peak associated with the Potts transition has moved up in temperature and is located at $T_3$ that we extract from the specific heat (see Fig.~\ref{fig:figure_Delta_1}). The presence of a sharp peak, which persists up to the largest system sizes we consider, $L = 300$, is a clear indication of a Potts transition at $\Delta = 1.0$. Note that since $B_E \propto L^{-2}$ we have rescaled the $L=300$ curve by $(300/80)^2$ to lie on top of the $L=80$ curve. In contrast, such a sharp peak is notably absent at $\Delta = 1.5$, where one only observes a change in slope around $T = 1.6 J$ associated with the nematic KT transition $T_2$.
    }
    \label{fig:figure_binder_mag_energy}
\end{figure}

The inset of Fig.~\ref{fig:figure_small_Delta}(b) schematically depicts the ordering that occurs at the lower transition: in the hexatic phase there exist different domains of the three-state Potts order parameter $n_i \in \mathbb{Z}_3$ (see Eq.~\eqref{eq:2.6-locking_constraint}). Regions with different $n_i$ are separated by domain walls that also represent domain wall strings for the nematic angle, where $\phi$ winds by $2 \pi/3$. There exist fractional vortices with $q_\phi = 1/3$ at the end of these strings.
In this regime, the cost of these domain walls is set by the nematic gradient energy, which is proportional to $J_2$, which explains why $T_3 \rightarrow 0$ as $\Delta \rightarrow 0$. The transition at $T_3$ corresponds to the ordering of the $\mathbb{Z}_3$ variable $n_i$, which occurs via a 2D Potts phase transition. Since $n_i$ is a discrete degree of freedom, the system exhibits true LR Potts order below $T_3$.
This scenario of an upper hexatic KT transition, where $\theta$ variables develop QLRO, and a lower Potts transition, where $n_i$ ($\phi$) develop LRO (QLRO, respectively) can also be observed in the Binder cumulants of the respective magnetizations, as shown in the left column of Fig.~\ref{fig:figure_binder_mag} in Appendix~\ref{appendix3}. It is also clearly seen in the Binder cumulant of the energy $B_E$, shown in Fig.~\ref{fig:figure_binder_mag_energy}. At the lower transition at $T_3$, the Binder cumulant $B_E$ develops a sharp peak, which is a clear indication of a second-order phase transition. In contrast, close to the upper transition $T_6$, $B_E$ only features a change in slope at about $1.1 \, T_6$, which is known to correspond to its behavior close to a KT transition. The Binder cumulant $B_E$ is, therefore, a convenient way to distinguish a second-order phase transition from a KT transition, as will be discussed more below.
Finally, we note that such an order of phase transitions (upper KT, lower Potts) has previously been reported for the generalized XY model~\cite{jiangMonteCarloSimulation1993,canovaKosterlitzthoulessPottsTransitions2014, de2016multiple} that corresponds to the $\lambda \rightarrow \infty$ limit of Eq.~\eqref{eq:2.1-model_phi}, as it is mentioned in section~\ref{ssec:previous_studies}.


\subsection{Upper Potts and lower KT transition at $0.9 < \Delta < 1.15$}
\label{sub:upper_Potts_middle_Delta}
Let us now discuss the region of main interest in the phase diagram around $\Delta = 1.0$. As illustrated by the Binder cumulants $B_\theta, B_\phi, B_\sigma$ in the middle column of Fig.~\ref{fig:figure_binder_mag} in Appendix~\ref{appendix3}, the ordering of hexatic, nematic and Potts degrees of freedom all occur at nearby temperatures in this region. As shown in Fig.~\ref{fig:figure_coupled_PD}, we find that the Potts (KT) transition temperature increases (decreases) monotonically with increasing $\Delta$ until the three Binder cumulants cross at a value of $\Delta = 0.9$. For $0.9 < \Delta < 1.15$, the Potts transition occurs above the KT transition. As the KT transition temperature exhibits a minimum at $\Delta = 0.9$, both transitions track each other with similar slope in that region. While the separation of the two transitions is small, $(T_3-T_{\text{KT}})/T_3 \approx 1\%$, we can resolve them within error bars using extensive Monte Carlo simulations up to system sizes of $L = 380$.

This is demonstrated in Fig.~\ref{fig:figure_Delta_1}, which shows results for $\Delta = 1.0$. Figure~\ref{fig:figure_Delta_1}(a) shows the spin stiffness $\rho$ and the specific heat $c$ that are used to extract KT and Potts transition temperatures, respectively. Specifically, we extract $T_3(L)$ from a fit of $c(T,L) \propto t^{-\alpha}$ with $t = [T-T_3(L)/T_3(L)]$ and Potts exponent $\alpha = 1/3$. As shown in detail in Fig.~\ref{fig:figure_Delta_1}(b), we determine the KT transition temperature $T_2$ from the universal jump criterion $\rho(T, L) = 2 T_2(L)/\pi$. We note that the nature of the KT transition changes from hexatic to nematic at $\Delta =0.9$.
Figure.~\ref{fig:figure_Delta_1}(c) displays the resulting system size dependent transition temperatures $T_2(L)$ and $T_3(L)$. We determine the transition temperatures in the thermodynamic limit by extrapolating to infinite system sizes using the expected scaling forms $T_2(L) - T_2(\infty) = a/(\ln L)^2$ and $T_3(L) - T_3(\infty) = a'L^{-1/\nu}$ with Potts correlation length exponent $\nu = 5/6$. We find that $T_3(\infty) > T_2(\infty)$ within a confidence of more than one standard deviation, specifically $T_3(\infty) = 1.2022 \pm 0.0005$ and $T_2(\infty) = 1.194\pm 0.002$ at $\Delta = 1.0$. This is one of the main results of this work. In the thin region $T_3 > T > T_2$ the system exhibits LR order in the relative hexatic-nematic orientation $n_i$ even though both angles are still only short-range correlated with a finite correlation length. We believe that this scenario of inverted Potts and KT transitions can explain the experimental observations in thin films of 54COOB, as discussed above.
\begin{figure}
    \centering
    \includegraphics[width=\linewidth]{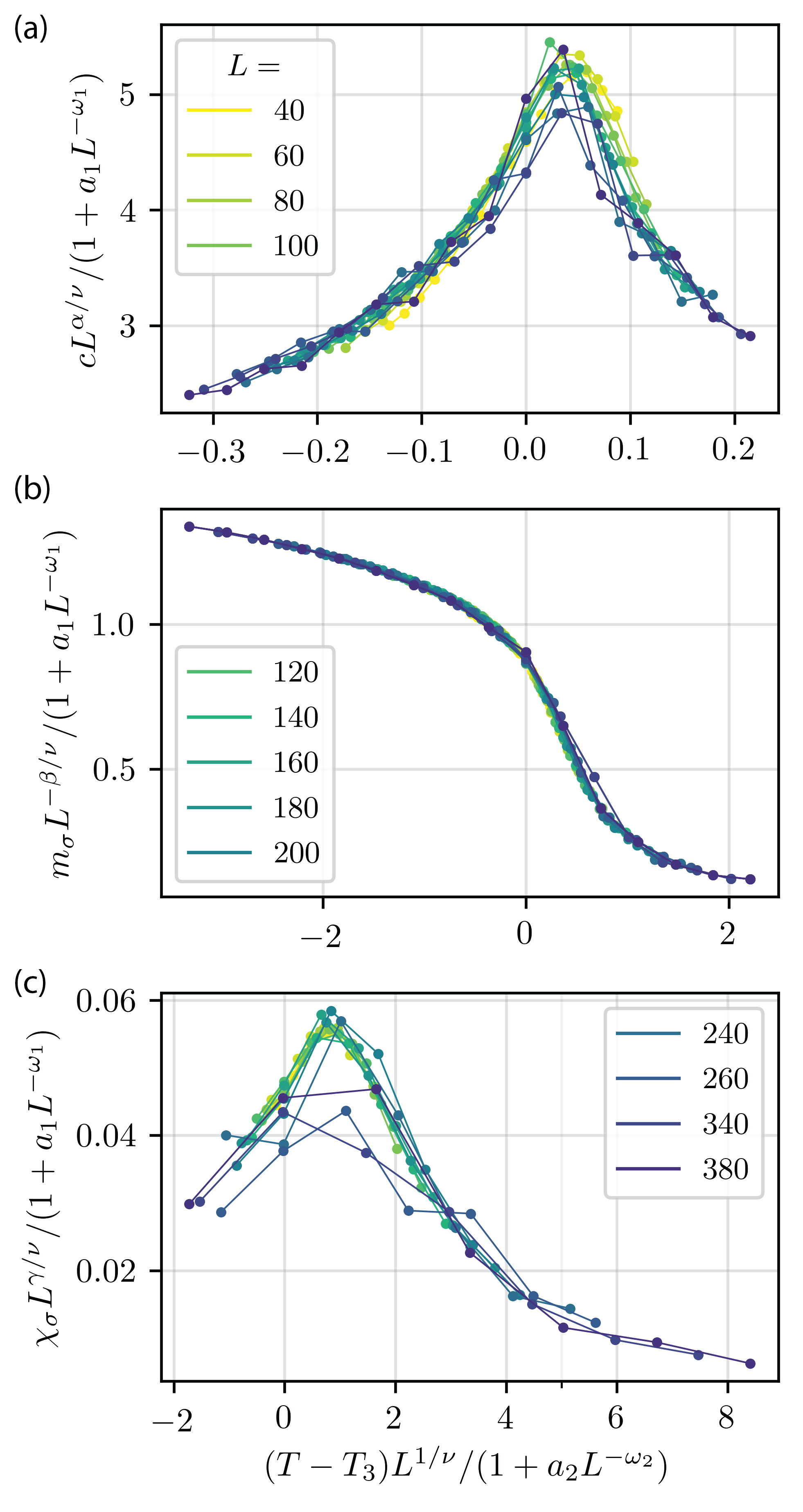}
    \caption{Finite size scaling analysis of (a) specific heat $c$, (b) Potts magnetization $m_\sigma$, and Potts susceptibility $\chi_\sigma$ for system sizes between $L = 40$ and $L=380$. We use the indicated scaling functions, which correspond to a 2D continuous Potts transition and include corrections to scaling. The resulting critical exponents and the Potts transition temperature $T_3$ are collected in Table~\ref{tab:scaling_exponent}. Data collapse is best for $m_\sigma$, but also works fairly well for $c$ and $\chi_\sigma$. }
    \label{fig:figure_scaling_analysis}
\end{figure}

Once the system enters a region with LR Potts order, the only available asymptotically free vortex excitations are nematic $q_\phi = 1$ vortices, formed of three bound elementary $q_\theta = 1$ vortices, confined by the Potts domain wall's finite tension. As discussed in Sec.~\ref{ssub:vortex_fractionalization_in_potts_ordered_phase}, the energy cost of these nematic vortices can be significantly lowered at finite values of $\lambda$ by expanding the vortex core to host three hexatic vortices of $q_\theta = 1$. This decreases the transition temperature for the unbinding of such extended combined vortices by a factor of $1/\ln[R/a]^2$, where $R > a$ is the vortex core size and $a$ is lattice scale. As a result, the KT transition temperature for unbinding of nematic vortices may be pushed below $T_3$, as we observe in our MC simulations. We emphasize that while we have focused on the value of $\lambda = 2.1$ in our work, we indeed find a region with $T_3 > T_2$ also for other values of $\lambda = 0.5, 1.0, 4.0$, demonstrating that our conclusions hold for an extended regime of hexatic-nematic couplings. We suggest that future work should explore the regime of small $\lambda$ in more detail, since the splitting of $T_3$ and $T_2$ is potentially larger there, and one could observe them directly.

Still, we note that we cannot fully exclude a scenario with a single transition in the Potts universality class, but we consider this unlikely at finite $\lambda$ based on our MC results. In other words, our numerical analysis shows that assuming that the stiffness makes a universal jump corresponding to an unbinding of nematic $q_\phi = \pm 1$ vortices, the associated KT transition would lie slightly below the Potts transition. In contrast, a single transition is expected in the $\lambda \rightarrow \infty$ limit of our model~\eqref{eq:2.1-model_phi}, in analogy to the phase diagram obtained for the vector-nematic generalized XY model version of our model~\cite{canova_competing_2016, shiBosonPairingUnusual2011,serna2017deconfinement}, where it is argued that vortex unbinding is suppressed by critical fluctuations of the discrete order parameter.
Further work, both numerical and analytical, should be directed towards the $\lambda \rightarrow \infty$ limit of Eq.~\eqref{eq:model-lambda-inf}, especially in the region of $\Delta \simeq 1.0$, to explore its phase diagram further and test this conjecture, especially in the region of $\Delta \simeq 1.0$. 

To further study the nature of the phase transitions, we perform a finite size scaling analysis of specific heat $c$, Potts magnetization $m_\sigma$, and susceptibility $\chi_\sigma$. As shown in Fig.~\ref{fig:figure_scaling_analysis}, a proper rescaling of the axes leads to a data collapse of results for different system sizes $40 \leq L \leq 200$ and temperatures. The collapse is most complete for the magnetization $m_\sigma$, but also evident for $c$ and $\chi_\sigma$. The critical exponents that we extract from the scaling analysis are consistent with a transition in the Potts universality class. The scaling analysis yields a transition temperature of $T_3 = 1.20 \pm 0.01$ that agrees with the more precise value that is obtained from the scaling of the maximum of the specific heat with $L$.
This confirms that the upper transition at $T_3$ lies in the Potts universality class, corresponding to the breaking of $\mathbb{Z}_3$ symmetry associated with the development of LR order in the relative orientation of hexatic and nematic angles.
\begin{table}
    \centering
    \begin{tabular}{|c|c|c|c|c|c|c|c|}
    \hline
    $\mathcal{O}$ & $T_c$ & $\zeta$ & $\nu$ & $a_1$& $\omega_1$ & $a_2$ & $\omega_2$ \\
    \hline
    \hline
    $c$ & $1.20$ & $\alpha = 0.30$ & $1.45$ & $10$ & $1.1$ & $100$ & $0.008$ \\
    \hline
    $m_\sigma$ & $1.20$ & $-\beta = -0.18$ & $1.25$ & $0.15$ & $0.13$ & $0.15$ & $0.19$\\
    \hline
    $\chi_\sigma$ & $1.20$ & $\gamma = 1.44$ & $0.95$ & $0.93$ & $0.11$ & $0.1$ & $0.05$\\
    \hline
    \end{tabular}
        \caption{Universal scaling exponents and critical temperature $T_c$ extracted from finite size scaling analysis of $c$, $m_\sigma$ and $\chi_\sigma$, shown in Fig.~\ref{fig:figure_scaling_analysis}. We use the scaling form $\tilde{\mathcal{O}}(t,L) = L^{\zeta/\nu} \mathcal{O}[L^{1/\nu} (T-T_c)/(1 + a_2 L^{-\omega_2}), L]/(1+a_1 L^{-\omega_1})$, which includes corrections to scaling. The 2D Potts critical exponents are $\alpha = 1/3$, $\beta = 1/9 \approx 0.11$, $\gamma = 13/9 \approx 1.44$, and $\nu = 5/6 \approx 0.83$, which is in fair agreement with our findings with the exception of $\nu$. Note that when tracking only $T_3$, and including small system sizes, we obtain a value of $\nu$ that lies much closer to its expected value (see Fig.~\ref{fig:figure_Delta_1} as well). }
    \label{tab:scaling_exponent}
\end{table}

\subsection{Single KT nematic transition at $\Delta > 1.15$}
\label{sub:single_KT_larger_Delta}
Finally, we briefly discuss the phase diagram at larger values of $1.15 \leq \Delta \leq 2$, where only a single nematic KT transition occurs as a function of $T$ (see Fig.~\ref{fig:figure_coupled_PD}). We identify the location and nature of the phase transition by the universal jump criterion of the stiffness, $\rho(T_2) = 2 T_2/\pi$, and find that the jump corresponds to an unbinding of nematic vortices of charge $q_\phi = 1$. Due to the locking constraint~\eqref{eq:2.6-locking_constraint}, these are equivalent to $q_\theta = 3$ hexatic vortices. Again, by fractionalizing the nematic vortices inside an extended vortex core so that there are three $q_\theta =1$ vortices, the energy of such a composite vortex can be lowered, especially at smaller values of $\lambda$. As $T_2 < T_\lambda$, QLRO of the nematic degrees of freedom immediately leads to QLRO of the hexatic variables, as they feel a uniaxial potential with a single minimum. Both $\text{O(2)}$ and $\mathbb{Z}_3$ symmetries are thus broken at the nematic KT transition for $\Delta > 1.15$. This is confirmed clearly by the Binder cumulants $B_\theta, B_\phi, B_\sigma$ at $\Delta = 1.5$, which are shown in the right column of Fig.~\ref{fig:figure_binder_mag} in appendix~\ref{appendix3}. Both $B_\phi$ and $B_\sigma$ undergo a sharp transition at $T_2$, while $T_6$ exhibits a smoother crossover towards the value of $2/3$ consistent with QLRO as temperature is lowered. The Binder cumulant of the energy $B_E$ only shows a change in slope above $T_2$ and does not exhibit any additional feature at temperatures $T < T_2$.

Finally, the specific heat exhibits two (broad) maxima, one at a temperature above $T_2$, which corresponds to the usual broad KT maximum at $1.1\, T_2$, where nematic vortices proliferate. In addition, there appears a maximum below $T_2$, that roughly follows the hexatic phase transition in the uncoupled system at $\lambda = 0$ (see Fig.~\ref{fig:figure_uncoupled_PD}). This crossover is due to the production of $q_\theta = 1$ hexatic vortices that are bound to fractionalized and confined $q_\phi = 1/3$ vortices. These hexatic vortices are thus bound together by domain wall strings of finite tension and only release a small part of the entropy, leading to a less pronounced maximum of $c$.


\section{Conclusion and outlook} \label{sec:conclusion}
Motivated by an long-standing experimental mystery in liquid crystal films of 54COOBC, we have presented a detailed Monte Carlo simulation study of  
a generalized XY model with finite coupling strength $\lambda$
between the two angular degrees of freedom.  
Our key finding is that there exists a 
parameter region in the phase diagram 
(for $\Delta \approx 1$ in Eq.~\ref{eq:2.1-model_phi}), where a 
sharp specific heat signature occurs at a temperature ($T_3$) above that
where the spin stiffness jumps to its finite value; all the thermodynamic exponents are consistent
with $\mathbb{Z}_3$ symmetry. This phase has relative nematic-hexatic order, and can be characterized as one of free $q_\phi=1$ nematic vortices formed of three bound $q_\theta=1$ hexatic vortices. At lower temperatures, our numerical
caling analysis demonstrates the presence of the Kosterlitz-Thouless transition ($T_{\text{KT,2}}$ or $T_{\text{KT, nem}}$), where composite nematic-hexatic vortices become bound. For the coupling value ($\lambda = 2.1$) in our simulations we find the ratio of temperature scales to be $T_3/T_{\text{KT},2} = 1.007$. Both nematic and hexatic degrees of freedom only develop quasi-long range order at $T_{\text{KT},2}$. Our simulations show no signs of first-order behavior anywhere in the phase diagram.

This scenario closely resembles the experimental observations in single-layer films of 54COOBC with a sharp specific heat signature 
between two smectic phases (Sm-A to Sm-A') at $T_{\text{Sm-A'}} = 66^\circ$C and a lower KT transition into a hexatic phase (Sm-A' to Hex-B) at $T_{\text{Hex-B}} = 63^\circ$C. The ratio of experimental transition temperatures is given by $T_{\text{Sm-A'}}/T_{\text{Hex-B}} = 1.018$, consistent with our findings. The transition into the Sm-A' phase is characterized by a sharp specific heat divergence with $\alpha \approx 0.3$, in agreement with the 2D Potts exponent $\alpha = 1/3$ that we obtained in our investigation of this minimal model. While experimentally the hexatic correlation length makes a sudden jump into the Sm-A' phase, it remains finite and hexatic QLR order only develops at $T_{\text{Hex-B}}$, again resembling our simulation results of a lower KT transition.

There are several additional features that emerge from our study. 
For example we reproduce previously reported results at smaller values of $\Delta$,
where a Potts transition \cite{jiangMonteCarloSimulation1993} 
occurs below a hexatic KT transition, 
$T_3 < T_{\text{KT},6}$ ($= T_{\rm KT, hex}$), that is characterized by the unbinding 
$q_\theta=1$ vortices.
Each hexatic defect is attached to a Potts domain wall 
(see Figure~\ref{fig:figure_intro2} (a)), so that the binding of these defects is accompanied
by the formation of a network of local Potts domains. The prior
establishment of a Potts transition at lower temperatures can be understood
as the vanishing of these Potts walls and the formation of a single
Potts domain. 

Our analytic estimates suggest that confinement fractionalization play
important roles in the emergence of the composite Potts phase.
At a temperature $T_{\rm conf} > T_{\text{KT},2}$, $q_\phi=1$ vortices
form composed of bound-states of three $q_\theta=1$ defects (see Figure \ref{fig:figure_intro2} (d)). Again the absence of ``dangling'' Potts walls means that a network
of local Potts domains is formed. Then the question is whether
the Potts ordering $T_3$ is above or at $T_{\text{KT},2}$ since there can be no
Potts domain walls once the $q_{\phi}$ defects bind.  
The extended vortex core sizes lower 
the KT transition to a value below $T_3$, revealing a new phase. 
Though our numerical studies yield
sharp thermodynamic features at a temperature $T_3$, 
it is not possible to distinguish a crossover from
a transition here since our system sizes are typically less than
the average separation between free defects. 

In order to maintain
continuity and consistency with previous computational studies, we explored
the hexatic-nematic XY model (Eq.~\ref{eq:2.1-model_phi})
with the same coupling value ($\lambda=2.1$) as
before \cite{jiangMonteCarloSimulation1993}. 
Since the vortex core size increases with decreasing 
hexatic-nematic coupling strength $\lambda$, we expect that the two temperature
scales $T_3$ and $T_{\text{KT},2}$ will be further separated at smaller 
values of $\lambda$, which should be explored in the future. Significantly larger system sizes 
should be accessible using techniques such as the worm algorithm \cite{prokof2001worm, shiBosonPairingUnusual2011, serna2017deconfinement} and matrix product states \cite{song2021hybrid}.

An alternative tuning parameter is hexatic ring-exchange coupling~\cite{saito1982melting,saito1982monte,strandburg1983monte,iaconis2012analytical, strandburgRevMod}, that is known to 
tune vortex core energies and could thus be used to suppress the lower 
KT transition, expanding the Potts phase. Tuning this parameter can be done numerically, and it has been shown experimentally that core energies and sizes are correlated with liquid crystal densities~\cite{strandburgRevMod}.  Adjusting these parameters may be done by varying local density, possibly by bending the
freely-suspended liquid crystal films of 54COOBC.  
Because the proposed Potts phase involves relative hexatic-nematic ordering
and hence higher-order correlations, it is challenging
to probe experimentally since most probes measure two-point correlation
functions.  

Taking our cue from intriguing results in a related
model~\cite{shiBosonPairingUnusual2011,serna2017deconfinement, roy2020quantum}, it is an
open question whether the two temperature scales $T_3$ and $T_{KT,2}$ 
converge into one single transition at large values of $\lambda$. The two end points in $\Delta$ of our composite Potts relative phase are worth further study. At $\Delta = 0.9$, the possible existence of a multicritical fixed point of $\text{O(2)} \times \mathbb{Z}_3$ symmetry suggests the exciting possibility of an emergent supersymmetry \cite{huijseemergent2015}, in analogy to the FFXY \cite{hasenbusch2005multicritical}. Close to $\Delta = 1.15$, where the Potts transition disappears, there may be a deconfined critical point~\cite{shiBosonPairingUnusual2011,serna2017deconfinement, roy2020quantum}. 

In conclusion we have presented the 54COOBC liquid crystal film problem as an 
experimental realization of the deconfinement of fractional vortices; here
the ``mystery'' phase above the nematic KT transition is one of composite
Potts ordering in the relative hexatic-nematic angles. We have reproduced the observed specific heat behavior as a function of temperature in our computational study of a minimalist model. The emergent Potts phase hosts extended nematic vortices that are analogous to baryons. Furthering the analogy, the three hexatic vortices that form such a bound state are akin to quarks. From this perspective, the experimentally observed Potts transition in the liquid crystal samples is a laboratory realization of quark confinement and baryon formation. 
Our work raises several further questions regarding the 
deconfinement of fractional vortices, in particular the need for a fully developed analytic theory for the interaction of domain walls with fractional vortices. We hope that our results will stimulate further interest in these problems.

\acknowledgments
We thank A. P. Young, P. Fendley, C. C. Huang, E. J. König, K. Chen, L. Jaubert, J. Schmalian, and L. Radzihovsky for useful discussions, and we are grateful to David Vanderbilt for 
providing us access to the Beowulf cluster at Rutgers University. This work was supported by the U.S. Department of Energy (DOE), Office of Science, Basic Energy Sciences under awards
DE-SC0020353 (P. Chandra) and DE-FG02-99ER45790 (P. Coleman and V.D.T.). V.D.T. also acknowledges the support of the Fonds de Recherche Qu\'{e}b\'{e}cois en Nature et Technologie. Part of the research (P.P.O.) was performed at the Ames Laboratory, which is operated for the U.S. DOE by Iowa State University under Contract DE-AC02-07CH11358.

\section*{Appendices}

\appendix

\section{Relevance of hexatic-nematic coupling} \label{appendix1}

We explore the relevance of the hexatic-nematic coupling $\propto \lambda \cos{[6(\tilde\vartheta - \tilde\varphi)]}$ in Eq.~\ref{eq:1.1-model_varphi}. The intermediate numerical factors are different for the models of Eqs.~\ref{eq:1.2-model_xy} and \ref{eq:2.1-model_phi}, but the end result is identical. We first turn to a long wavelength version of the model

\begin{equation}
\begin{split}
\tilde{\mathcal{H}}&= \int d^2 r \big[ \frac{K'_6}{2} (\nabla \tilde\vartheta)^2 + \frac{K'_2}{2} (\nabla \tilde\varphi)^2 \\
& - \lambda_3 \cos{[6(\tilde\vartheta - \tilde\varphi)]}\big]
\end{split}
\label{eq:appendix1}
\end{equation}

 For $K'_2 = K'_6$ (i.e. $\Delta = 1.0$ in Eq.~\ref{eq:2.1-model_phi}), one has that the hexatic $T_{\rm KT, 6}$ and nematic $T_{\rm KT, 2}$ are at identical temperatures, i.e. $T_{\rm KT, 6}=T_{\rm KT, 2}$. We can find whether the coupling constant $\lambda_3 = \tilde{\lambda}/T$ is a relevant perturbation at the KT transition temperature by using the fact that at the transition $T_{\rm KT, p}$ the exact value of the renormalized stiffnesses is known to be 
\begin{align}
  \label{eq:app1a}
  K'_{R,p}(T_{\rm KT, p}) = \frac{2 p^2}{\pi} \,,
\end{align}
\noindent with bare $K'_p = J_p/T$, from the model presented in Eq.~\ref{eq:1.1-model_varphi}. Following Refs. \cite{kadanoff1990scaling, chaikin1995principles}, we evaluate the correlation function associated with the coupling between $\vartheta$ and $\varphi$, assuming no correlations between $\tilde\vartheta$ and $\tilde\varphi$, \emph{i.e.}, $\av{[ \tilde\vartheta(x) - \tilde\varphi(0)]^2} = 0$, valid for $\lambda_3 = 0$, via calculating 

 \begin{align}
   \mathcal{C}_{\lambda} (x) &= \av{ e^{6 i [\tilde\vartheta(x) - \tilde\vartheta(0) - \tilde\varphi(x) + \tilde\varphi(0) ]}} \nonumber \\
   & = e^{ - \frac{36}{2} \av{[\tilde\vartheta(x) - \tilde\vartheta(0)]^2}  - \frac{36}{2} \av{[\tilde\varphi(x) - \tilde\varphi(0)]^2}}  \nonumber   \\
&= e^{- \frac{36}{2 \pi} \bigl( \frac{1}{K'_{R,6}} + \frac{1}{K'_{R,2}} \bigr) \ln \frac{x}{a_0}} \nonumber\\
&= \left| \frac{x}{a_0} \right|^{-\eta_c}\,, \label{eq:app1b}
 \end{align}
 
\noindent where we used $ \frac12 \av{[\psi(x) - \psi(0)]^2} \approx \frac{1}{2 \pi K} \ln \left| \frac{x}{a_0} \right|$ for Gaussian field $\psi$ with coupling $K$. The exponent is
\begin{align}
  \label{eq:app1c}
  \eta_c = \frac{18}{\pi} \Bigl( \frac{1}{K'_{R,6}} + \frac{1}{K'_{R,2}} \Bigr) \,.
\end{align}

Via the Kadanoff construction  \cite{kadanoff1990scaling, chaikin1995principles}, we find whether $\lambda_3$ is a relevant perturbation by determining the sign of the scaling dimension

\begin{align}
  \label{eq:app1d}
  \mathcal{D}_{\lambda_3} = 2 - \frac{\eta_c}{2} = 2 - \frac{9}{\pi} \Bigl( \frac{1}{K'_{R,6}} + \frac{1}{K'_{R,2}} \Bigr) \,.
\end{align}

Using Eq.~\eqref{eq:app1a} at $T_{\rm KT, 6}=T_{\rm KT, 2}$, we obtain
\begin{align}
  \label{eq:app1e}
  \mathcal{D}_{\lambda_3} \bigr|_{T = T_{\rm KT, 6}=T_{\rm KT, 2}} = \frac34 > 0 \,.
\end{align}
This means that $\lambda_3$ is a \emph{relevant} perturbation at $T =T_{\rm KT, 6}=T_{\rm KT, 2}$ and the system will flow away from the KT transition at non-zero $\lambda_3$, and hence the system will encounter a high-temperature $T_{\lambda}$ scale where the variables become locked with one another, satisfying the $\lambda$ coupling. This completes the derivation of the result presented at equation~\ref{eq:2.4}, obtained for model \ref{eq:2.1-model_phi} but otherwise identical.

\section{Estimation of the size of composite vortices.} \label{appendix2}

In this section, we provide a rough estimate of the size of a composite vortex of total charge $q_{\phi} = 1$ formed of three hexatic vortices ($q_{\theta}=1$) bound through a domain wall of the $\sigma$ variable, such that $\sigma \rightarrow \sigma + 2\pi/3$ across such wall. This procedure was done in the following references \cite{babaevphase2004,goryo2007deconfinement, nitta2012baryonic}, with relevance to quark deconfinement, and in \cite{radzihovsky2008superfluidity} for the $\mathbb{Z}_2$ symmetric case. We largely reproduce the treatment of this last reference here for our $\mathbb{Z}_3$ symmetric case.

We consider an effective long wavelength model analogous to equation~\ref{eq:2.1-model_phi}, obtained by transforming $\cos{(\theta_i - \theta_j)} \sim dr (\nabla \theta)^2/2$ into the continuum limit, with $K_2 = \Delta/T$, $K_6 = (2 -\Delta)/T$ and $h = \lambda/T$:

\begin{equation}
\begin{split}
\bar{H}&= \int d^2 r \big[ \frac{K_6}{2} (\nabla \theta)^2 + \frac{K_2}{2} (\nabla \phi)^2 \\
& - h \cos{(\theta - 3 \phi)}\big]
\end{split}
\label{eq:appendix1}
\end{equation}

Note that one can relate $K_2$ and $K_6$ to those of the original model of Eq.~\ref{eq:1.1-model_varphi}, used in appendix~\ref{appendix1}, such that $K_p = p^2 K'_p$. We consider two types of point defects in the two XY variables

\begin{equation}
\oint_{r_0} \nabla \theta \cdot d\bm{r} = 2\pi q_{\theta} \qquad \oint_{r_0} \nabla \phi \cdot d\bm{r} = 2\pi q_{\phi}
\end{equation}

\begin{figure}
    \centering
    \includegraphics[width=\linewidth]{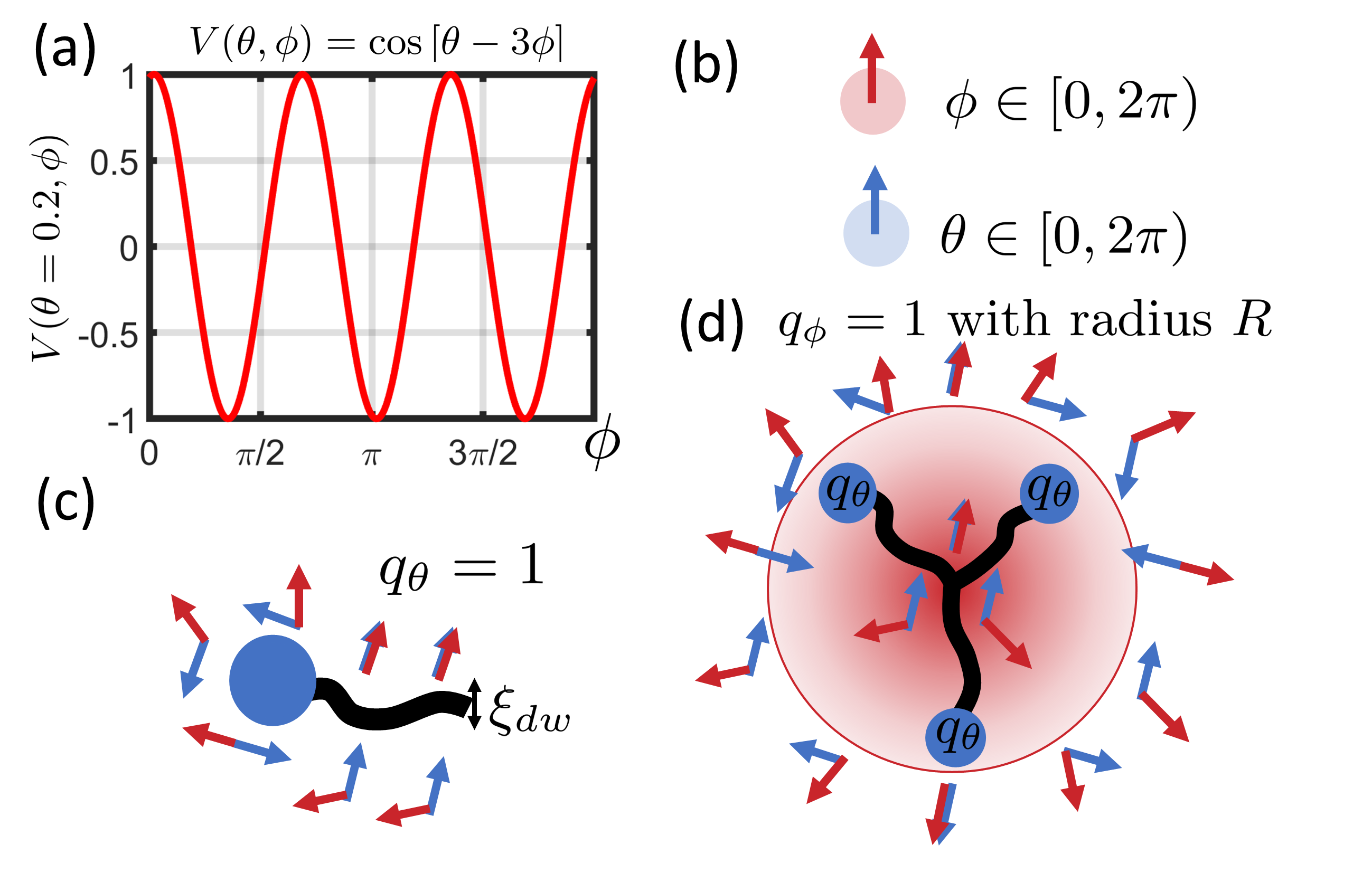}
    \caption{(a) Illustration of the $\lambda$ constraint; domain walls of the relative variable $\sigma = 2\pi n/3$ (as defined in Eq.~\ref{eq:def_sigma}), which can be of width $\xi_{dw}$, are situations where the $\theta$ and $\phi$ do not fall in the deep minima of $V$. (b) The domain of both $\theta$ and $\phi$, corresponding to the model of Eq.~\ref{eq:2.1-model_phi}. This is different than the $\vartheta$ and $\varphi$ variables from Fig.~\ref{fig:figure_intro2}. Here, both are $\text{O(2)}$ variables. (c) Visualization of a hexatic vortex $q_{\theta} = 1$, with its adjacent domain wall of the relative variable $\sigma = 2\pi n/3$ (as defined in Eq.~\ref{eq:def_sigma}), of width $\xi_{dw}$. This can also be viewed as a fractional $q_\phi = \frac13$ vortex. (d) A composite vortex of total charge $q_{\phi} = 1$ and $q_{\theta} = 3$, which is clearly seen outside the core. It is formed of three fractional ones bound together by their domain wall strings. These are arranged in this simple geometry where the walls are of length $R$ and the distance between $q_{\theta} = 1$ vortices is $\sqrt{3}R$. 
    }
    \label{fig:figure_appA}
\end{figure}

It is clear that the two variables have their own independent behavior for $h = 0$. However, for any finite $h$, an extensive energy cost is incurred if the system does not lock the two phases on average with each other such that $\av{\theta} = 3 \av{\phi}$. This further leads to the winding numbers to be related to each other, as it is explained in Eq.~\ref{qlambda3}. The first way to satisfy this is to simply have $q_{\phi} = 1$ and $q_{\theta} = 3$. However, as we will show later, this is very costly as the phase has to wind very tightly around the vortex, generating a large core energy for the vortex. The second way to satisfy this constraint is to have $q_{\theta} = 1$ and $\Delta n = 1$ (or equivalently, $q_\phi = 1/3$) In the simplest picture, this fractional vortex in the $\phi$ generates a line defect originating at the vortex core (see Fig.~\ref{fig:figure_appA}), through which $\phi$ will rapidly wind by $2\pi/3$. It is possible however that some part of the rapidly winding ``leaks'' into the $\theta$ variable. For $h$ extremely large or $K_i$ very small, we expect that this wall defect will be very thin ($\xi_{dw} \ll \xi_i$), but as $h$ is decreased, it should become wider and wider, leaving to $\phi$ more space to wind the remaining. Note that in both case, the energy of such a wall will scale linearly with the length of the wall, such that $E_{dw} \propto L$. 

These walls separate domains of different $n = -1, 0, 1$ where we have $2\pi n/3 = \sigma = \frac13(\theta - 3 \phi)$, following Eq.~\ref{eq:def_sigma}. One can clearly isolate a solution for the domain wall by rewriting the phase variables
\begin{align}
& \alpha = \frac{\theta + 3\phi}{2}   \qquad   \sigma = \frac{\theta - 3\phi}{3} \\
\text{or} \qquad  & \theta = \alpha + \frac{3\sigma}{2} \qquad \phi = \frac{\alpha}{3} - \frac{\sigma}{2}
\end{align}
This transformation is valid as long as the phases are well-defined. The model of equation~\ref{eq:appendix1} is rewritten as
\begin{equation}
\begin{split}
\bar{H}&= \int d^2 r \big[ \frac{K_a}{2} (\nabla \alpha)^2 + \frac{K_b}{2} (\nabla \sigma)^2 \\
&- K_{ab} (\nabla \alpha ) \cdot (\nabla \sigma)   - h \cos{(3\sigma)}\big] 
\end{split}
\end{equation}
with
\begin{align}
K_a &= (K_6 + K_2/9) \\
K_b &=(9 K_6 + K_2)/4 \\
K_{ab} &= \frac{1}{2}(K_2/3 - 3K_6 )  
\end{align}
Understanding that this Hamiltonian is simply $\bar{H} = \int d^2 r \mathcal{E}(\alpha(r), \sigma(r))$ with $\mathcal{E}$ the enery density, we write the saddle point equations $\delta \mathcal{E}/\delta \alpha = \delta \mathcal{E}/\delta \sigma = 0$ which leads to:
\begin{align}
-K_a \nabla^2 \alpha + K_{ab} \nabla^2 \sigma &= 0\\
-K_b \nabla^2 \sigma + K_{ab} \nabla^2 \alpha + 3h \sin{(3\sigma)} &= 0
\end{align}
We can eliminate $\alpha$ from these equations and find the know sine-Gordon equation for the $\sigma$ variable.
\begin{equation}
\begin{split}
&-K\nabla^2 \sigma + 3h \sin{(3\sigma)} = 0 \\
&\qquad \text{with} \qquad K = K_b - K_{ab}^2/K_a
\end{split}
\end{equation}
We solve this equation for the situation where we have $\sigma(x,y \rightarrow \infty) \rightarrow 0$ and $\sigma(x,y \rightarrow -\infty) \rightarrow 2\pi /3$ (this is done without loss of generality, as all domain walls have a jump of $2\pi/3$), which leads to a domain wall along the $x$ direction. The solution for this domain wall is then simply
\begin{equation}
\begin{split}
&\sigma_{dw} (y) = \frac{4}{3} \arctan{(e^{2y/\xi_{dw}})} \\
&\qquad \text{with} \qquad \xi_{dw} =\frac{2}{3}\sqrt{\frac{2K}{3h}}
\end{split}
\end{equation}
We get that $\alpha_{dw}(y) = \frac{K_{ab}}{K_a} \sigma_{dw} (y)$ and then, for our initial variables

\begin{equation}
\begin{split}
\theta_{dw} &= \left( \frac{3}{2} + \frac{K_{ab}}{K_a} \right) \sigma_{dw} (y) \\
\phi_{dw} &= \frac{1}{2} \left( \frac{2K_{ab}}{3K_a} - 1 \right) \sigma_{dw} (y)
\end{split}
\end{equation}
One has that the energy of a domain wall of length $L$ is $E_{dw} = \epsilon_{dw} L$ with
\begin{align}
\epsilon_{dw} &= \int \text{d}y \big[ \frac{K_a}{2} (\nabla \alpha_{dw})^2 + \frac{K_b}{2} (\nabla \sigma_{dw})^2 \\
& \; - K_{ab} (\nabla \alpha_{dw} ) \cdot (\nabla \sigma_{dw})   - h \cos{(3\sigma_{dw})} \big] \\
&= \int \text{d}y \big[ \frac{K}{2} (\nabla \sigma_{dw})^2  - h \cos{(3\sigma_{dw})} \big] \\
&= \int \text{d}y \big[ \frac{K}{2} (\nabla \sigma_{dw})^2  + \frac{K}{2} (\nabla \sigma_{dw})^2 \big] \\
&=  \int \text{d}y K (\nabla \sigma_{dw})^2 \\
&= \frac{16 K}{9 \xi_{dw}} = 4 \sqrt{\frac{2K h}{3}} \label{eq:app_edw}
\end{align}
There is a caveat to this result. As $K \rightarrow 0$ then the energy of the wall goes to $0$ and so does its width. On the other hand, as $h \rightarrow \infty$, the wall width goes to $0$ but the energy increases dramatically. It is important in this context to bring back the lattice cutoff $a \sim 1$ here. Hence, we have $\xi_{dw} = \frac{2}{3}\sqrt{\frac{2K}{3h}} > 1$. For $h < 4K/9$ the theory is well behaved. For larger $h$ one has to use $\xi_{dw} = 1$ and adjust Eq.~\ref{eq:app_edw}.  Note that putting back the initial parameters $K_2$, $K_6$ and $h$  in the domain wall energy we get $K = 9(K_2 K_6) /(K_2 + 9 K_6)$ which leads to the following expression for the domain wall energy:
\begin{equation}
\epsilon_{dw} =  4 \sqrt{6 h \left(\frac{K_2 K_6 }{K_2 + 9 K_6}\right) }  \label{eq:dwener}
\end{equation}
or, using $\Delta$, 
\begin{equation}
\epsilon_{dw} =  4 \sqrt{3 h \left(\frac{\Delta (2-\Delta) }{9 - 4 \Delta}\right) }  \label{eq:dwener2}
\end{equation}

\begin{widetext}

With this new insight, we wish to compare the energy of a point defect with $q_\phi = 1$ to the assembly of three hexatic objects with each $q_\theta = 1$, linked by a domain wall, for the same total charge. This arrangement is shown in Fig.~\ref{fig:figure_appA}. We denote the first situation as a point vortex (pv) and the second as a split vortex (sv). The energy of a pure defect with no domain wall, i.e. one in which a $q_{\theta} =3$ vortex is accompanied by a $q_{\phi} = 1$ vortex, is simply expressed as:

\begin{align}
E_{\rm pv} &= E_{c,\theta} (3) + E_{c,\phi} (1) + 9\pi K_6 \ln{(L/\xi_6)} + \pi K_2 \ln{(L/\xi_2)} \\
&= \big[ \frac{\pi^2}{2} ( 9 K_6 + K_2) + 9\pi K_6 \ln{(R/\xi_6)} + \pi K_2 \ln{(R/\xi_2)} \big] + 9\pi K_6 \ln{(L/R)} + \pi K_2 \ln{(L/R)} \\
&= E_{\rm pv}^c (R) + 9\pi K_6 \ln{(L/R)} + \pi K_2 \ln{(L/R)} \label{eq:pv_ene}
\end{align}

the second line uses the first approximation to the core energy of the vortices $E_c \sim \pi^2 q^2 K/2$, with $q$ being the charge of the vortex. Note that in this formula, we have explicity put is an arbitrary radius $R$, and split the log parts. This will come in handy when we compare this new core energy $E_{\rm pv}^c (R)$ to the split core energy $E_{\rm sv}^c (R)$, as we consider $R$ to be the radial length of a split combination of vortices. We will then be able to compare only the energetics inside the radius $R$, since for $r > R$, both combinations will look and act like a point vortex. This point vortex is thought to be the dominant kind as $h \gg K$, since then the domain wall width becomes extremely small and its energy very large.

In the case of a split vortex, we need to include some extra energetic terms, i.e. the logarithmic repulsion between vortices themselves $V_{q-q} (r) = - 2\pi K q^2 \ln{(r/\xi)}$. There are three such terms here for each type of vortex, and they are separated by a distance of $\sqrt{3}R$ for this simple geometry (see Fig.~\ref{fig:figure_appA}). We also include the attractive domain wall energy $\epsilon_{dw} R$ for each of the domain walls. We then get

\begin{align}
E_{\rm sv}^c (R) &= 3(E_{c,\theta} (1) + E_{c,\phi} (1/3) ) - 3 \times 2\pi K_6 \ln{(\sqrt{3}R/\xi_6)} - 3\times \frac{2\pi}{9} K_2 \ln{(\sqrt{3}R/\xi_2)} + 3 R \epsilon_{dw} \\
&= \frac{3\pi^2}{2}(K_6  + K_2/9 ) - 6 \pi K_6 \ln{(\sqrt{3}R/\xi_6)} - \frac{2\pi}{3} K_2 \ln{(\sqrt{3}R/\xi_2)} + 3R \epsilon_{dw} \label{eq:sv_ene}
\end{align}

\end{widetext}

At this point, we stress that point configurations, the point vortex and the split vortex, are topologically equivalent at $r > R$. We can then compare the two configurations in order to determine the region in which a split vortex would be less energetic than the point vortex. 

We start by finding out the optimal size of a split vortex by finding $R_{\rm max}$ such that $\frac{\partial  E_{\rm sv}^c (R)}{\partial R} \rvert_{R_{\rm max}}= 0$. In the following, we used $\xi_6 = \xi_2 = a$. This leads to the following expression for the optimal radial size of a split vortex

\begin{equation}
R_{\rm max}/a = 2\pi \frac{K_2/9 + K_6}{\sqrt{3}\epsilon_{dw}}
\end{equation}

One sees that driving the coupling $h$ to infinity completely annihilates the split vortices structures (by $\epsilon_{dw} \rightarrow 0$). There is however a regime with low $h$ where split vortex will be quite extended. Note that this expression does not depend on the temperature, but only on the interaction parameters themselves.

In the case of strong coupling, i.e. $h > 4K/9$ (for $K_2 = K_6 = 1$, this is $h > 4/10$) then the domain wall width is $\xi_{dw} = a = 1$ i.e. the lattice spacing. This is the case for the $\lambda = 2.1$ we have chosen for our simulations. In this case, we get

\begin{equation}
\sigma_{dw} (y) = \frac{4}{3} \arctan{(e^{2y})}
\end{equation}

which then leads to 

\begin{align}
\epsilon_{dw} &=  16 K/9
\end{align}

and then, for $\Delta = 1.0$, our computational value, we have

\begin{align}
R_{\rm max}/a \simeq 2.5
\end{align}

which completes the derivation of the vortex size estimate that is presented in the text, in section~\ref{ssub:vortex_fractionalization_in_potts_ordered_phase}. Using this value for the radius of the composite vortex into the energy estimates from equations~\ref{eq:pv_ene} and \ref{eq:sv_ene}, we get that for such an extended vortex,

\begin{equation}
\begin{split}
E_{\rm pv}^c (R_{\rm max})  &\simeq  78.13 \\
E_{\rm sv}^c (R_{\rm max}) &\simeq 5.98
\end{split}
\end{equation}

proving that in the regime of our $\mathbb{Z}_3$ relative order phase, composite vortices are favored with respect to point ones due to their extended nature, even with the cost of domain walls in the core of the vortex.


\section{Comparison of the Binder cumulant for different regimes.} \label{appendix3}

\begin{figure*}
    \centering
    \includegraphics[width=.9\linewidth]{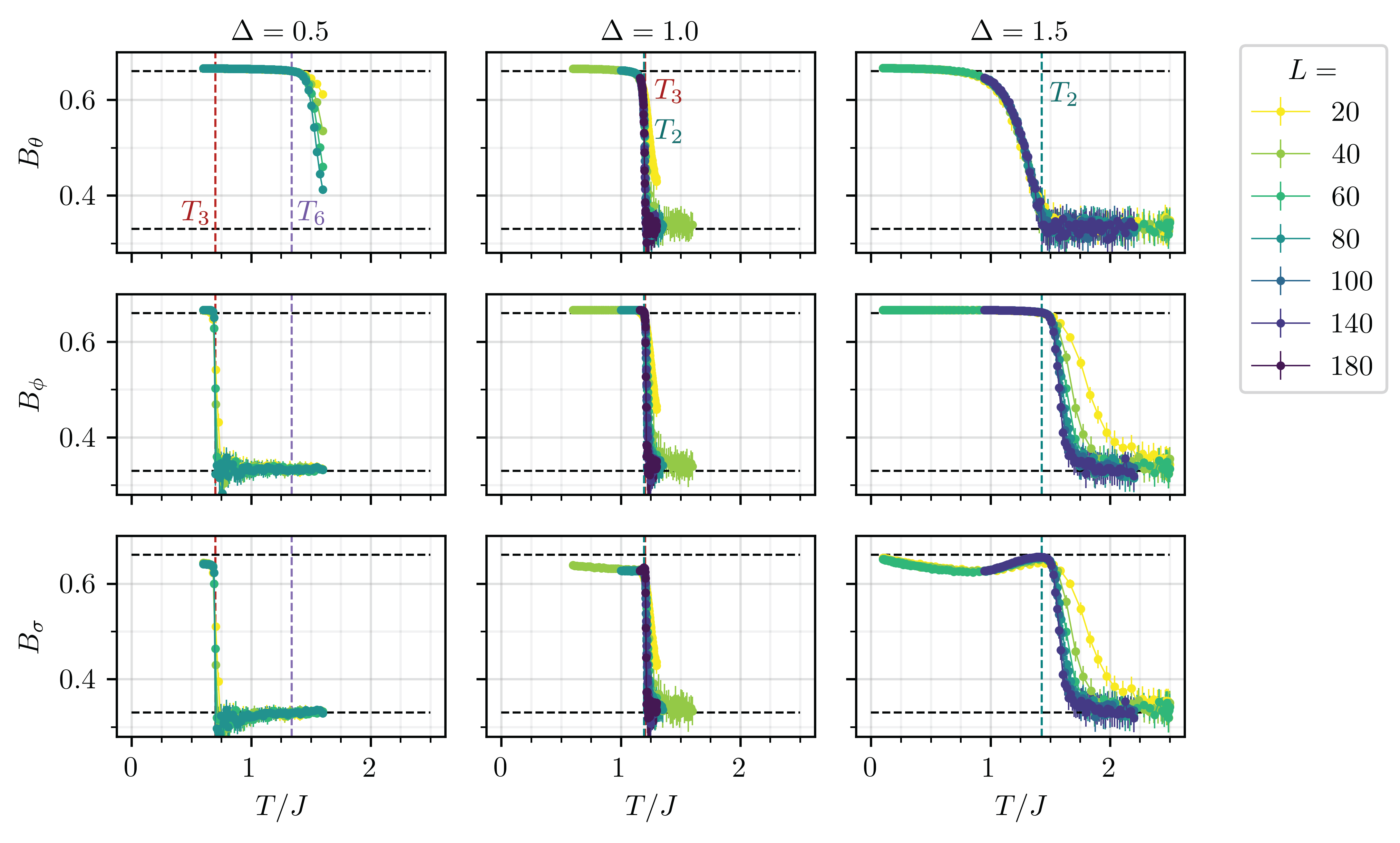}
    \caption{Binder cumulants $B_\theta, B_\phi, B_\sigma$ of hexatic $\theta$, nematic $\phi$ and relative $\sigma = (\theta - 3 \phi)/3$ degrees of freedom as a function of temperature $T/J$. Different columns denote different values of $\Delta = 0.5, 1.0, 1.5$ as indicated. Vertical lines indicate location of transition temperatures $T_3$ (Potts), $T_6$ (KT hexatic), and $T_2$ (KT nematic). 
    }
    \label{fig:figure_binder_mag}
\end{figure*}

We present here a comparison of the Binder cumulant for $M_a$ with $a = \theta, \phi, \sigma$, computed from equation~\ref{eq:3.3b}. This is shown in Fig.~\ref{fig:figure_binder_mag}. The Binder cumulants approach a value of $1/3$ $(2/3)$ in the disordered (ordered) state at high (low) temperature. In the $\Delta = 0.5$ column, one observes that hexatic and Potts phase transitions are clearly separated with $T_6 > T_3$, as signaled by the rapid increase of the respective Binder cumulants. At the Potts transition, both nematic and Potts Binder cumulants rapidly increase and the system is fully (algebraically) ordered below $T_3$. Note that the increase of $B_\theta$ is signifcantly more broadened than that of $B_\phi$ and $B_\sigma$ as expected for a KT transition. In contrast, at $\Delta = 1.5$, we observe that the nematic and Potts Binder cumulants increase at a higher temperature than the hexatic one. The transition is broad showing that this is a KT transition. Once the nematic degree of freedom is ordered, it induces a potential for the hexatic, which then tends align with the nematic as temperature is lowered, similarly to a spin in a magnetic field. Correspondingly, the increase of $B_\theta$ occurs over a rather broad temperature range, corresponding to a crossover into the fully ordered state. Finally, at $\Delta = 1.0$ we observe that all three Binder cumulants sharply increase close to $T = 1.2 J$, suggesting the presence of a Potts transition. The crossing $B_\sigma(L)$ yields a transition temperature $T_3 \simeq 1.202$ that is consistent with our findings in Fig.~\ref{fig:figure_coupled_PD}. Note that the Binder cumulants do not allow us to easily address the question whether the nematic and Potts transitions are separated. To demonstrate this we rather rely on a refined scaling analysis, presented in Fig.~\ref{fig:figure_Delta_1} in the main text.


\bibliography{EPotts_Biblio}

\end{document}